\documentclass[12pt]{iopart}
\usepackage{graphics}
\usepackage{array}

\begin{document}

\title{Dirac electronic states in graphene systems:\\ Optical spectroscopy studies}

\author{M. Orlita$^{\dag}$ and M. Potemski}

\address{Laboratoire National des Champs Magn\'{e}tiques Intenses,
CNRS-UJF-UPS-INSA,\\
25, avenue des Martyrs, 38042 Grenoble, France}
\ead{milan.orlita@grenoble.cnrs.fr,
marek.potemski@grenoble.cnrs.fr}

\begin{abstract}
Electronic properties of two-dimensional allotropes of carbon,
such as graphene and its bilayer, multi-layer epitaxial graphene,
few-layer Bernal-stacked graphene, as well as of three-dimensional
bulk graphite are reviewed from the viewpoint of recent optical
spectroscopy studies. Attention is focused on relativistic-like
character of quasi particles in these systems, which are referred
to as massless or massive Dirac fermions.
\end{abstract}

\pacs{81.05.Uw,82.45.Mp,73.20.-r,78.20.-e} \maketitle

\section{Introduction}
The family of carbon-based materials, including two distinct crystal forms: the
isotropic diamond and anisotropic graphite, together with other allotropes such
as fullerenes and carbon nanotubes, has been recently enlarged to include
graphene, a unique material which consists of a two-dimensional (2D) lattice of
carbon atoms with a honeycomb symmetry. Graphene, the closest archetype of a
two- dimensional crystal, while being very promising for future applications,
represents also an extremely interesting system from the viewpoint of
fundamental physics~\cite{GeimNatureMaterial07,CastroNetoRMP09}. Distinctly,
the electronic states of graphene do not obey the conventional laws of
Schr\"{o}dinger's quantum mechanics but are rather governed by an equation
which is equivalent to the Dirac equation for massless
fermions~\cite{NovoselovNature05,ZhangNature05}. This very peculiar, but at the
same time simple, electronic band structure of graphene has been investigated
theoretically for more then sixty years~\cite{WallacePR47}. The ``rise'' of
graphene dates however only from 2004 when it was effectively introduced into
the the laboratory environment, following the clear identification of ~10
$\mu$m sized graphene flakes deposited on Si/SiO$_2$
substrates~\cite{NovoselovScience04} and of multi-layer epitaxial graphene
thermally decomposed from silicon carbide~\cite{BergerJPCB04,BergerScience06}.
More recently, the growth of large area graphene samples on metallic surfaces
(Ru, Ni or
Cu)~\cite{SutterNM08,PargaPRL08,GruneisPRB08,KimNature09,GruneisNJP09,LiScience09},
and their subsequent transfer on arbitrary substrates has also substantially
progressed~\cite{KimNature09,LiScience09}, what may be a promising option for
easy preparation of macroscopic-size crystallites of graphene.

The intense, over last years, investigations of different graphene
systems have expanded today into a broad research field of
qualitatively new two-dimensional systems~\cite{GeimScience09}.
The challenging expectation are that graphene will introduce
quantum electrodynamics into solid state
laboratories~\cite{KatnelsonNaturePhys06,KatsnelsonEPJB06,BeenakkerRMP08,YoungNaturePhys09}
and that it will constitute the basis for future electronics
plausible avoiding the known limitations (at the nano-scale) of
the CMOS technology. As for today, the basic properties of ``new''
2D allotropes of carbon, including
graphene~\cite{NovoselovNature05,ZhangNature05}, graphene
bilayer~\cite{NovoselovNaturePhys06,McCannPRL06}, multi-layer
graphene~\cite{NovoselovScience04,CraciunNaturNano09}, graphene on a silicon
carbide substrate~\cite{BergerJPCB04}, are already relatively
well-known and the basis of the graphene physics is pretty well
established.

Despite relevant advances in various experimental techniques, such
as angular resolved photoemission spectroscopy
(ARPES)~\cite{OhtaScience06,BostwickNaturePhys07,ZhouNatureMater07,GruneisPRL08}
or scanning tunnelling spectroscopy
(STS)~\cite{MartinNaturePhysics07,LiPRL09II,MillerScience09}, a
large part of our knowledge on electronic properties of graphene
has been deduced from conventional, electric transport and optical
experiments, in particular under an applied magnetic field. The
physics of the integer quantum Hall effects in single-layer and
bilayer graphene
\cite{NovoselovNature05,ZhangNature05,NovoselovNaturePhys06} or of
the minimum conductivity of graphene
\cite{TanPRL07,ChenNaturePhys08} is nowadays underpinned by a
solid basis of published transport experiments. Optical
investigations of graphene systems are equally fertile. They offer
a rather direct verification of electronic band structure in these
materials and/or help to avoid the advanced sample processing of
often invasive character~\cite{BlakeSSC09}.

A number of papers which review the graphene-related physics has
already appeared within last few
years~\cite{GeimPhysToday07,GeimNatureMaterial07,GusyninIJMP07,CastroNetoRMP09,KatsnelsonMT07,KatsnelsonSSC07,GeimScience09},
but the progress in this research area is extremely fast. This, in
particular, concerns the developments related to optical
spectroscopy which was perhaps not very popular at the birth of
the research on graphene but it is now more and more frequently
used, as testified by a large number of recent papers cited here.
As any review article also this one includes some unavoidable
repetitions of information which can be found elsewhere. However,
our intention is to discuss the properties of different graphene
systems from the viewpoint of optical spectroscopy methods,
including the most recent developments in this field. Whereas the
previous review articles were mostly focused on the properties of
a single graphene monolayer, here we discuss the Dirac-like
electronic states (massless and massive) which in fact are
representative of any two-dimensional allotropes of carbon or even
of bulk graphite.

Indeed, the methods of optical spectroscopy have been recently
applied to a large class of graphene-based systems. A number of
various optical and magneto-optical measurements have been
performed on mono-, bi- and multi-layer Bernal-stacked graphene
(exfoliated from graphite and placed on Si/SiO$_2$), on multilayer
epitaxial graphene thermally decomposed from C-terminated surface
of SiC, as well as on bulk graphite. The published papers provide
a mosaic of information as they usually address a particular
problem or individual sample characteristics. Investigations of
the electronic band structure remains, however, the key element of
the majority of these papers and also constitutes the bulk of our
review. Other characteristic problems invoked with optical methods
concern the efficiency or mechanisms of carrier scattering. The
effects of disorder or interactions are also here discussed.

The paper is organized as follows. The next section has still some
introductory character in which we stress the relevance of the
optical spectroscopy in the ``rise'' of graphene, i.e., in the
identification of the graphene flakes on Si/SiO$_{2}$ substrates.
The following sections 3--6 are systematically devoted to: 3)
monolayer graphene and multilayer epitaxial graphene (on C-face of
SiC) which both exhibit the simple Dirac-like electronic spectrum,
4) bilayer graphene with the unique massive Dirac-like fermions 5)
bulk graphite, with more complex electronic bands but also
displaying the Dirac-like dispersion relations, and 6) $N$-layer
($N>2$) segments of graphite. In each of these section we first
present the theoretically expected band structure and optical
response of the investigated system, then we confront it with the
available experimental data and discuss the implications. The
particular attention is focused on magneto-optical investigations,
which we consider to be particularly valuable in the studies of
electronic properties of the investigated systems. Conclusions and
possible perspectives related to optical studies of graphene
systems are presented in Section 7.

\section{How to see graphene}

\begin{figure}
\begin{center}
\scalebox{0.35}{\includegraphics*{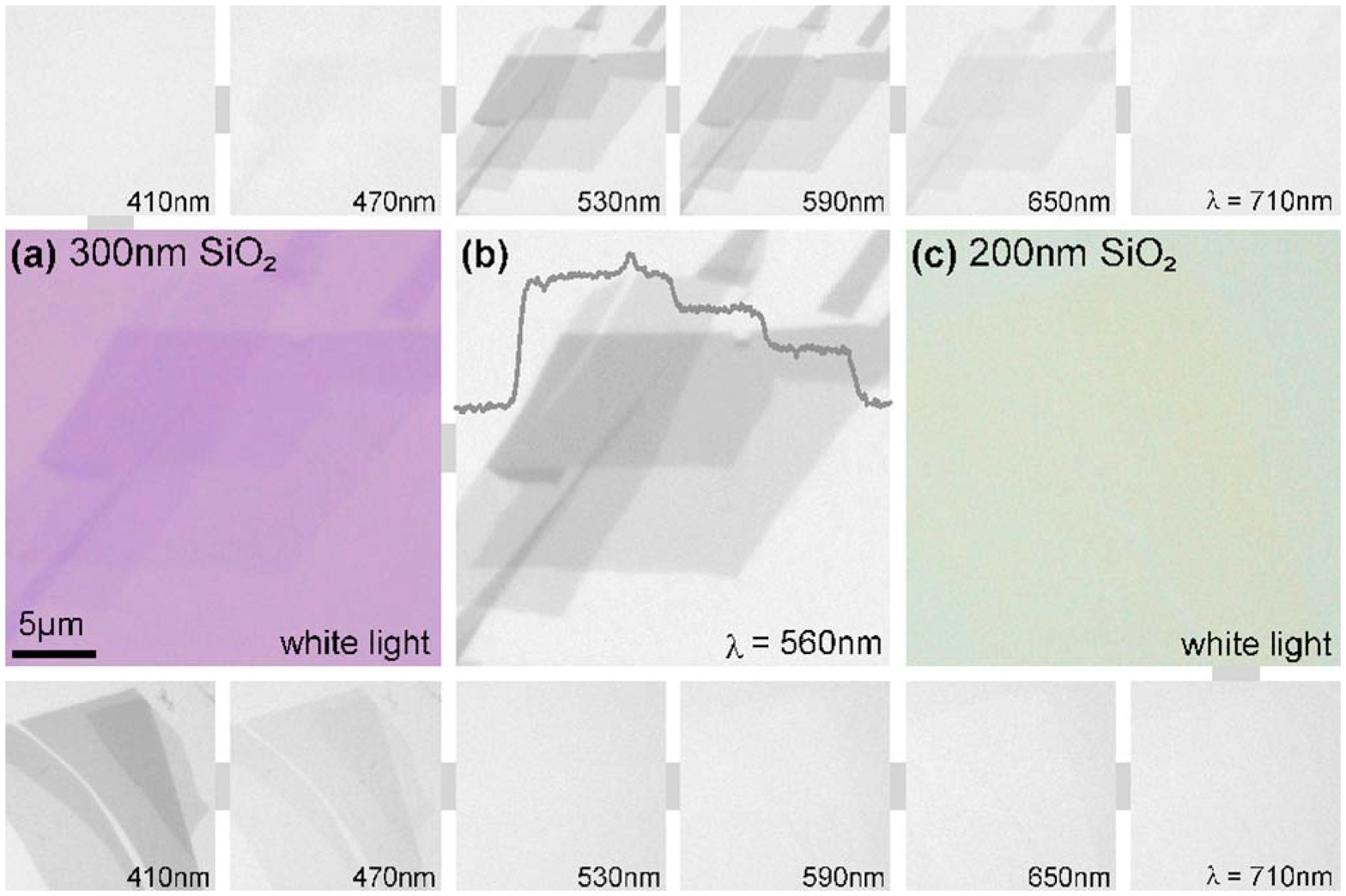}} \hspace{0.1cm}
\scalebox{0.40}{\includegraphics*{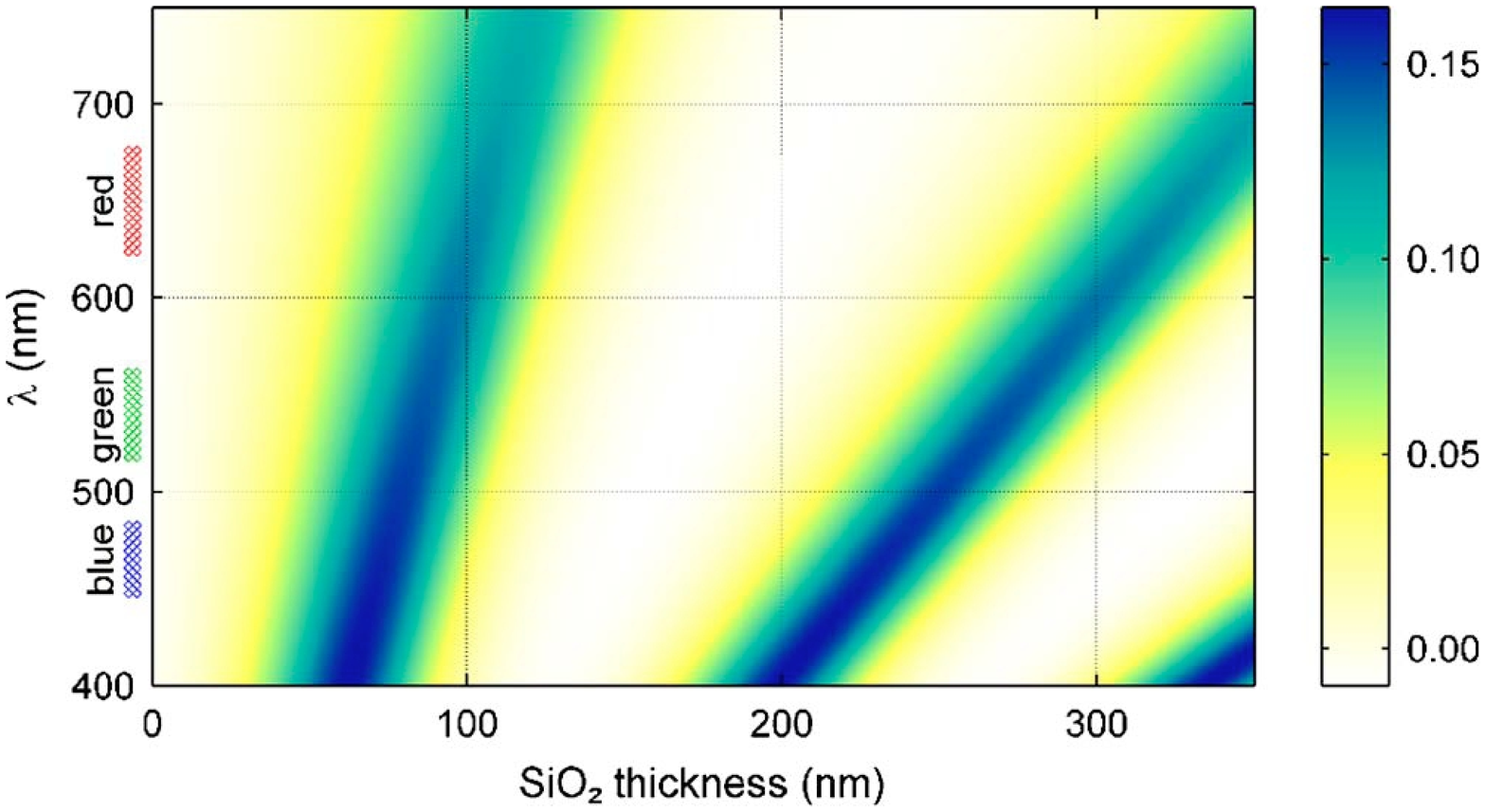}}
\end{center}
\caption{\label{Contrast} Graphene crystallites on the Si/SiO$_2$
substrate with the oxide thickness of 300 nm imaged in white light
(a), green light (b) and another graphene sample on the Si
substrate with 200-nm-thick SiO$_2$ layer imaged in white light
(c). The smaller figures in the upper and lower lines represent
images taken at various wavelengths for the thickness of oxide
layer 300 and 200 nm, respectively. The theoretically calculated
contrast (for graphene monolayer) as a function of the wavelength
and the  thickness of the oxide layer is shown in part (d). In a
relatively good agreement with these calculations, graphene
crystallites are best visible in images taken at wavelengths
slightly below 600~nm and around 400 nm, for the thickness of
dioxide layer 300 and 200 nm, respectively. Reprinted with
permission from \cite{BlakeAPL07}. Copyright 2007, American
Institute of Physics.}
\end{figure}

The identification of the ``exfoliated graphene'' on Si/SiO$_2$
substrates by the group at the Manchester
University~\cite{NovoselovScience04} is one of the unquestionable
milestone in the development of the graphene oriented research.
Initially, however, it was perhaps not so clear why a few- or even
single-graphene sheets can be seen just through an optical
microscope. The explanation of this today widely accepted and
frequently used experience is not trivial and invokes the
importance of the spectroscopic analysis already at the birth of
the graphene physics. As more recently revealed in
details~\cite{BlakeAPL07,AbergelAPL07,RoddaroNL07,CasiraghiNL07,JungNL07},
the choice of the appropriate thickness of the SiO$_2$ layer plays
the essential role in the visualization of the graphene layers. As
shown in Fig.~\ref{Contrast}, the contrast which is necessary to
``see'' the individual graphene layers depends on the wavelength
of light and the actual thickness of the SiO$_2$ layer. This is
accounted by the analysis of multireflections from the sandwiched
graphene/SiO$_2$/Si structure within the framework of standard
Fresnel equations (despite the fact that we are dealing with the
material at the atomic resolution). The substrates used by
Manchester group were characterized by 300-nm-thick oxide layers,
for which the optimum contrast to see the graphene flakes is
$\sim$10~\% and falls in the middle of the visible range. This
greatly helped to visualize the graphene flakes.

\section{Graphene}
\label{MONO}

\subsection{Electronic band structure}

\begin{figure}
\begin{center}
\scalebox{0.4}{\includegraphics*{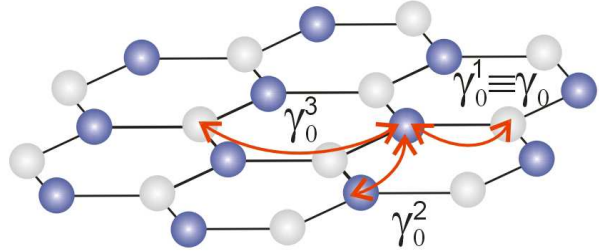}}
\end{center}
\caption{\label{Monolayer_atomic}  Segment of the graphene
crystal: Carbon atoms are arranged into a 2D lattice with a
characteristic honeycomb symmetry and lattice constant
$a_0=0.246$~nm \cite{ChungJMS02}, which is by a factor of
$\sqrt{3}$ larger than inter-atomic distance.}
\end{figure}

\begin{figure}
\begin{center}
\scalebox{0.5}{\includegraphics*{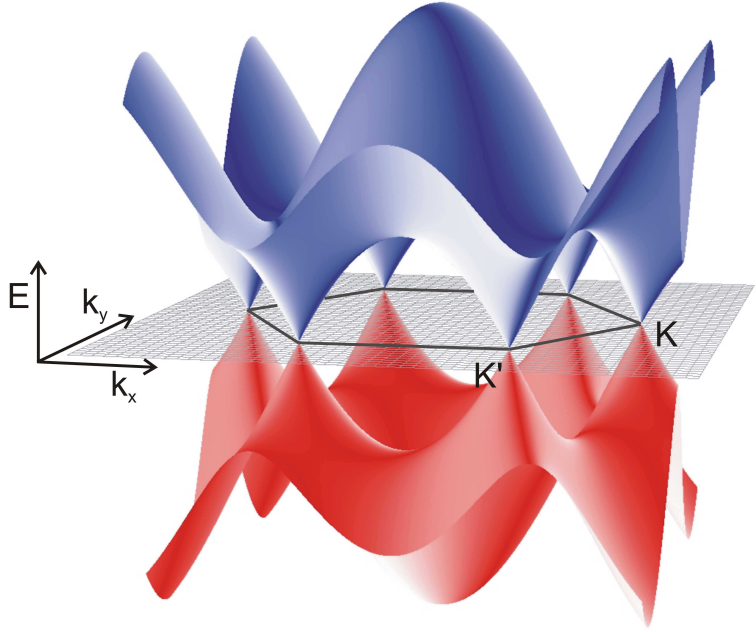}}
\end{center}
\caption{\label{Band_structure} Schematic view of the graphene
band structure with the characteristic Dirac cones in the vicinity
of $K$ and $K'$ points.}
\end{figure}

The first calculation of electronic states in a 2D lattice of
carbon atoms with a honeycomb symmetry, see
Fig.~\ref{Monolayer_atomic}, dates back to 1947, when
Wallace~\cite{WallacePR47} used graphene as a starting element for
description of bands in bulk graphite. Taking into account the
strong hybridization of $2sp^2$ orbitals in the graphene plane,
Wallace considered just the remaining $p$ orbital (oriented
perpendicular to the crystal plane) to be responsible for the
electronic band structure in the vicinity of the Fermi level and
suggested a standard tight-binding approach. Considering only the
nearest-neighbour hopping parameter $\gamma_0$ (see
Fig.~\ref{Monolayer_atomic}), one easily finds a pair of
$\pi$-bands \cite{CastroNetoRMP09}:
\begin{eqnarray}\label{Cosines}
 \lefteqn{E_{\pi^\ast}(\mathbf{k})=-E_{\pi}(\mathbf{k})=}\nonumber \\
\hspace{1cm}
=\gamma_0\sqrt{1+4\cos^2\left(\frac{k_ya_0}{2}\right)+4\cos\left(\frac{k_x\sqrt{3}a_0}{2}\right)\cos\left(\frac{k_ya_0}{2}\right)},
\end{eqnarray}
which distinctly cross (touch) at two inequivalent $K$ and $K'$
points of the Brillouin zone, see Fig.~\ref{Band_structure}. The
strength of the nearest-neighbour hopping is
$\gamma_0\approx3.2$~eV~\cite{Brandt88} and the lattice constant
$a_0=0.246$~nm~\cite{ChungJMS02} is by a factor of $\sqrt{3}$
larger than the distance between the nearest carbon atoms.

In pristine graphene, the Fermi level lies just at the touching
(crossing) point of $\pi$ and $\pi^{*}$ bands and graphene has a
character of zero-band-gap semiconductor (semimetal). Close to a
given crossing (touching) point, known also as the Dirac or charge neutrality
point, the electronic bands are nearly linear and practically
rotationally symmetric. In other words, the carrier dispersion
relations take a simple form: $E_{\pi^\ast}=-E_{\pi}\approx
v_F\hbar|\mathbf{k}|$, where the momentum $\mathbf{k}$ is measured
with respect to $K$~($K'$) point. The parameter $v_F$, having
dimension of a velocity, is directly related to the coupling
strength (hopping integral) between the nearest carbon atoms:
$v_F=\sqrt{3}a_0\gamma_0/(2\hbar)$~\cite{CastroNetoRMP09}.

The linearity of bands in graphene (in the vicinity of the $K$ and
$K'$ points) implies that charge carriers in this material behave
as relativistic particles with zero rest mass and constant
velocity $v_F\approx10^6$~m.s$^{-1}$. They are often referred to
as~\emph{massless Dirac fermions}, and with a good precision, their
behaviour is described by the effective
Hamiltonian~\cite{CastroNetoRMP09,PeresPRB06}:
\begin{equation}\label{Monolayer}
\hat{H}=v_F  \left(\begin{array}{cc}
0 & p_x-ip_y \\
p_x+ip_y &  0 \\ \end{array} \right)= v_F \left(\begin{array}{cc}
0 & \pi^\dag \\
\pi &  0 \\ \end{array} \right)= v_F
\mbox{\boldmath$\sigma$}\cdot\mathbf{p},
\end{equation}
which is equivalent to the Hamiltonian in the Weyl equation
for real relativistic particles with zero rest mass
(originally for neutrinos) derived from the Dirac equation. Due to this formal similarity, a
direct link between quantum electrodynamics and the physics of
graphene is established.

The characteristic linear dispersion relations of electronic
states makes graphene very distinct among other 2D systems (such
as quantum wells or heterojunctions) which have been widely
investigated in condensed-matter physics for last thirty years.
For instance, the density of states in graphene is not constant as
for conventional massive particles, $D=g_vg_s
m/\left(2\pi\hbar^2\right)$, but rises linearly with the energy
distance $\varepsilon$ from the Dirac point:
$D(\varepsilon)=g_sg_v |\varepsilon|/(2\pi v_F^2\hbar^2)$. Here,
$g_s=g_v=2$ stand for the spin and valley degeneracies,
respectively.

Nevertheless, it should be noticed that the relativistic-like
image of electronic states in graphene given by
Hamiltonian~(\ref{Monolayer}) remains an approximate model. This
simple approximation is well, and perhaps even surprisingly well
fulfilled in case of electronic states in the vicinity of the
Dirac point. Those states indeed show almost perfect character of
massless Dirac fermions when probed with different experiments.
However, as can be seen from Fig.~\ref{Band_structure}, the
deviations from this relativistic model become obviously important
for states far away from the Dirac point, even if we consider only
the nearest neighbours in the tight-binding calculation. Other
complications may arise when including the hopping integrals
between further located neighbours. For example, when taking into
account the non-zero values of $\gamma_0^{2,3}$ hopping
integrals~\cite{GruneisPRB08II,StauberPRB08} (see
Fig.~\ref{Monolayer_atomic}), the nonlinearity is enhanced and
Dirac cones become asymmetric with respect to the charge
neutrality point~\cite{PlochockaPRL08,StauberPRB08,DeaconPRB07}.

\begin{figure}
\begin{center}
\scalebox{0.65}{\includegraphics*{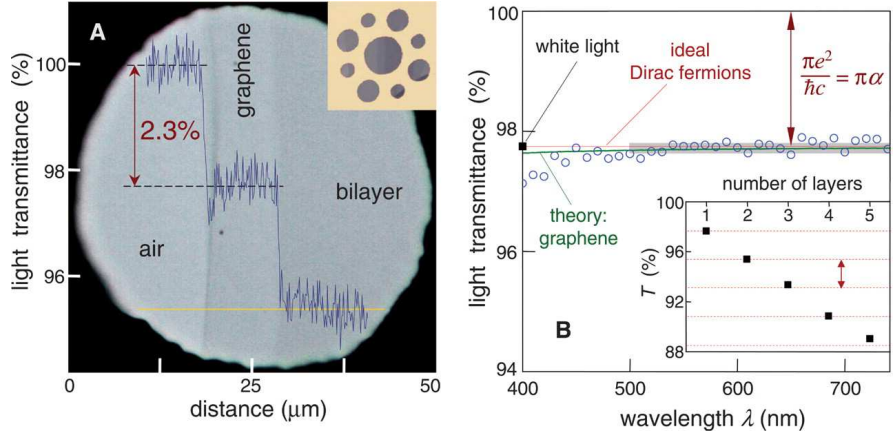}}
\end{center}
\caption{\label{Nair} (a) Photograph of a 50-mm aperture partially
covered by graphene and its bilayer, with schematically depicted
optical absorption. (b) Transmittance spectrum of single-layer
graphene in comparison with theoretical transmission for ideal
massless Dirac fermions and real graphene. Inset shows relative absorption as a function of number of layers in
graphite stacks. Reprinted from~\cite{NairScience08} with
permission from AAAS.}
\end{figure}

\subsection{Optical conductivity at $B=0$}

Interaction of massless Dirac fermions in graphene with light,
expressed usually in terms of dynamical (optical) conductivity,
has been theoretically treated at different levels of
approximation~\cite{AndoJPSJ02,GruneisPRB03,PeresPRB06,GusyninPRL06,MikhailovPRL07,MishchenkoPRL07,FalkovskyEPJB07,FalkovskyPRB07,StauberPRB08,StauberPRB08II,GrushinPRB09,SheehyPRB09}.
Within the simplest approach of electronic states described by
Hamiltonian~(\ref{Monolayer}), only vertical in $k$-space
transitions across the Dirac point are optically active. Then, the
dynamical conductivity takes a constant (energy independent) value
of $G_0=e^2/(4\hbar)$, which results in frequency-independent
absorption of $\pi\alpha\approx2.3$~\%, where $\alpha=e^{2}/\hbar
c$ (in cgs units) is the fine structure constant. Hence, in this
model, the absorption of light in graphene is so-called universal,
given just by $\alpha$ which in general describes the strength of
interaction between light and matter. Let us note that the
graphene dynamical conductivity related to the vertical across the
Dirac point transitions is frequency independent,
$\sigma(\omega)\propto(1/\omega)\overline{D}(\omega)=\mathrm{const}$,
simply because the joint density of states $\overline{D}(\omega)$
is linear with energy (frequency). When considering the linearly
polarized light, the contribution of individual $k$-conserving
transitions to the total optical conductivity depends on the angle
between the given momentum $\mathbf{k}$ and the vector of light
polarization~\cite{GruneisPRB03}, but the universal conductivity
is restored after integration over all angles.

Transforming $G_0$ into measurable quantities, such as the optical
transmission $T$ and reflection $R$, one finds that $R \ll T$
($R=\pi^2\alpha^2T/4$~\cite{StauberPRB08}), and therefore, the absorption of
light in graphene, and also the fine structure constant $\alpha$ itself, can be
estimated in a straightforward transmission experiment, $T\cong1-\pi\alpha$.
This fact was confirmed in recent experiments in the visible range on the
self-standing graphene membrane~\cite{NairScience08}, see Fig.~\ref{Nair}, and
further explored by other groups~\cite{MakPRL08,FeiPRB08}. The universality of
the optical conductivity in graphene has been also discussed in the context of
epitaxial graphene on SiC substrate~\cite{DawlatyAPL08II} in a wide range of
photon energies.

\begin{figure}
\begin{center}
\scalebox{0.8}{\includegraphics*{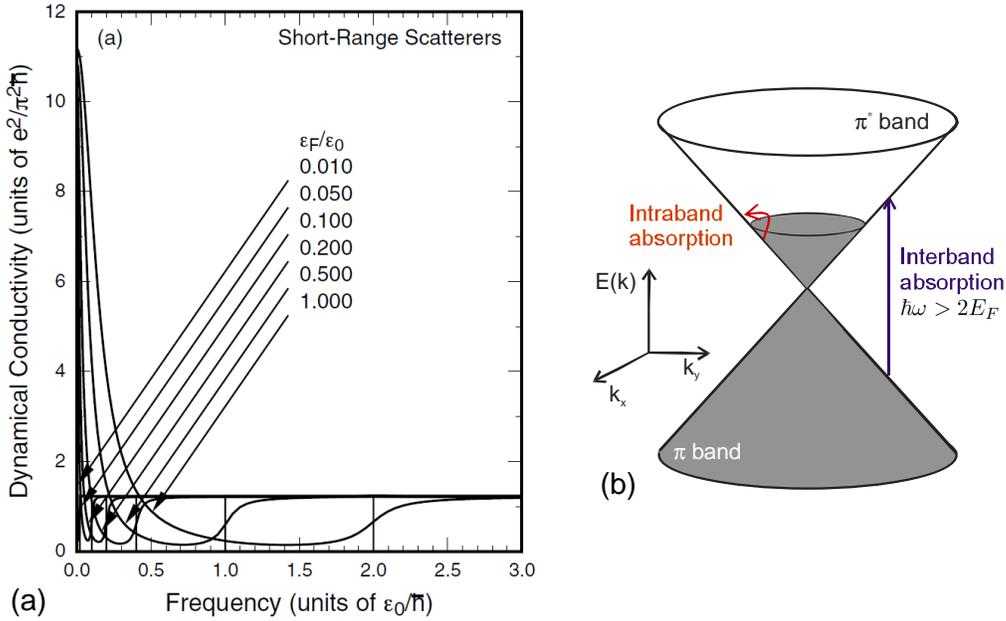}}
\end{center}
\caption{\label{Monolayer_transitions} (a) Real part of the
optical conductivity (per one valley) as a function of the Fermi
level (in units of an arbitrary energy $\varepsilon_0$) calculated
using the self-consistent Born approximation for short-range
scattering centers~\cite{AndoJPSJ02}, reprinted with permission of the Journal of the Physical Society of
Japan. (b) Schematic diagram of intra- and
inter-band transitions in doped graphene. The non-vanishing Fermi
level gives rise to the characteristic onset of optical
conductivity at $2E_F$. Intra-band transitions results in the
appearance of the Drude-peak centered at zero frequency.}
\end{figure}

Nevertheless, we should mention that the declared universality of
the graphene absorption is only approximative (what unfortunately
limits its applications in metrology), since
Hamiltonian~(\ref{Monolayer}) represents just the simplest model
of the graphene band structure and therefore neglects the
non-linearity of bands at high energy distance from the Dirac
point. In other words, the absorption of light in graphene is as
``universal'' as ``ideal'' are massless Dirac fermions in this
system. A theoretical analysis of these corrections was performed
by Stauber \emph{et al.}~\cite{StauberPRB08}, who discussed not
only deviations from non-linearity due to finite width of
$\pi$-bands as given by Eq.~(\ref{Cosines}), but also included the
effects of next-nearest hopping integral $\gamma_0^2$, see Fig.~\ref{Monolayer_atomic}.
Experimentally, the deviations of the optical conductivity from
the ``universal'' value $G_0$, have been probed in the visible
range by Fei and coworkers~\cite{FeiPRB08}. Another departure from
``universality'' of the graphene transmission is provided by saturation
effects,
due to which graphene can find its use as a saturable absorber~\cite{TanAPL10,BaoAFM09,ZhangOE09,MishchenkoPRL09}.

It is worth noticing that a nearly universal conductivity has been
also reported by Kuzmenko~\emph{et al.}~\cite{KuzmenkoPRL08} for
bulk graphite (conductivity per single atomic sheet). Naively, one
can seek for the origin of this effect in a relatively weak
inter-layer coupling in graphite, but a deeper theoretical
analysis is also available: the appearance of the universal
conductivity (again per sheet) in multi-layer graphene stacks has
been considered by Min and MacDonald~\cite{MinPRL9} in relation to
the presence of the chiral symmetry in structures with AB and/or
ABC stacking sequences.

\begin{figure}
\begin{center}
\scalebox{0.50}{\includegraphics*{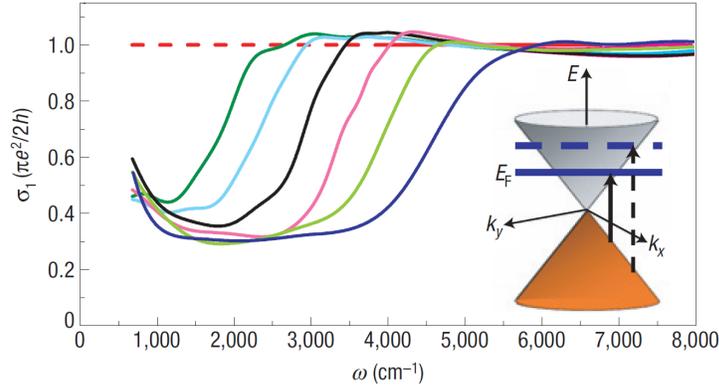}}
\end{center}
\caption{\label{Ando+Basov} The experimentally determined (real
part of) optical conductivity of graphene placed on Si/SiO$_2$
substrate for different positions of the Fermi
level~\cite{LiNaturePhys08}. Reprinted by permission from
Macmillan Publishers Ltd:  Nature Physics~\cite{LiNaturePhys08},
copyright (2008).}
\end{figure}

A more general theoretical approach to the optical conductivity of
graphene, which includes the effects of carrier density,
temperature, as well as carrier scattering (both short- and
long-range scatterers considered), has been presented already in
``pre-graphene'' era by Ando \emph{et al.}~\cite{AndoJPSJ02} in
the framework of the self-consistent Born approximation, see
Fig.~\ref{Monolayer_transitions}a. Later on, further calculations
were performed by Peres \emph{et al.}~\cite{PeresPRB06} as well as
Stauber~\emph{et al.}~\cite{StauberPRB08II}. A simple analytical
expression for the optical conductivity was found by
Gusynin~\emph{et al.}~\cite{GusyninPRL06} followed by Falkovsky
and Varlamov~\cite{FalkovskyEPJB07}. These works account for both
inter-band transitions which are blocked at energies below twice
the Fermi energy ($2E_F$) as well as intra-band transitions which
result in the appearance of the Drude peak centered at zero
frequency (see Fig.~\ref{Monolayer_transitions}). As can be
expected, these models uncover the universal conductivity
$G_0=e^2/(4\hbar)$ in the limit of high energies. If the Fermi energy
approaches the neutrality point, the optical conductivity is
universally  $G_0=e^2/(4\hbar)$ at any finite frequency but
singular at $\omega=0$. The apparent value of $\sigma(\omega =0)$
related to the so-called minimum conductivity is one of the lively
discussed issue in the graphene physics, without clear consensus
between different experiments and theoretical models so
far~\cite{AndoJPSJ02,NovoselovNature05,ZhangNature05,TanPRL07,AdamPNAS07,CheianovPRL07,ChenNaturePhys08}.

Experimentally, the optical response of exfoliated graphene flakes
placed on Si/SiO$_2$ substrates has been investigated in the
mid-infrared range in the reflection configuration by Wang
\emph{et al.}~\cite{WangScience08} who reported the pronounced
modification of graphene's optical properties as a function of the
position of the Fermi level. Further experimental data taken in
both reflection and transmission mode, accompanied by a detailed
analysis have been presented by Li~\emph{et
al.}~\cite{LiNaturePhys08} and later on also by Mak \emph{et
al.}~\cite{MakPRL08}. The spectrum of the optical conductivity
extracted from these experiments (see Fig.~\ref{Ando+Basov}) is in overall agreement with
theoretical predictions. However, some of the observed features,
such as the striking scaling of the onset of inter-band
transitions with the gate voltage or the intriguing finite
absorption in between the Drude peak and the $2E_{F}$ onset remain
to be firmly clarified. Possibly, these observations might be due
to effects of disorder and/or of electron-electron interactions,
which as yet may not be fully accounted in theoretical
models~\cite{MikhailovPRL07}.

\subsection{Magneto-spectroscopy}

\begin{table}[b]
\caption{\label{Table_graphene_quantities} Table of various graphene-related quantities together with their
numerical values. The energy $\varepsilon$ and the Fermi level $E_F$ in the last column are expressed in meV
and the Fermi velocity $v_F$ in units of $10^6$~m.s$^{-1}$. $m_0$ denotes the
bare electron mass.}
\begin{indented}
\item[]
\begin{tabular}{@{}llrl}
\br
Quantity &Relations& Numerical values& \\
\br
Effective mass & $m=|\varepsilon|/v_F^2$ & $1.76\times10^{-4}(|\varepsilon|/v_F^2)$&$m_0$\\
Density of states  &      $g_s g_v |\varepsilon|/(2\pi
v_F^2\hbar^2)$ & $1.47\times10^8(|\varepsilon|/v_F^2)$ & cm$^{-2}$.meV$^{-1}$\\
Carrier density & $g_s g_v E_F^2/(4\pi
v_F^2\hbar^2)$ &  $7.35\times10^7(E_F/v_F)^2$ & cm$^{-2}$ \\
Energy of 1$^{\mathrm{st}}$ LL  & $E_1=v_F\sqrt{2\hbar e B}$ & $36.3\, v_F  \sqrt{B[T]}$ & meV\\
\br
\end{tabular}
\end{indented}
\end{table}

Cyclotron motion of charge carriers and the related cyclotron
resonance phenomenon (absorption of light at cyclotron frequency
$\omega_C$) is primarily a classical effect, probably the most
representative for the magneto-optical spectroscopy. Such motion
is not only characteristic for a conventional charged ($e$)
particle with mass $m$, which precesses with the frequency of
$\omega_C=eB/m$. The solution of the classical equation of motion
for charged particle with energy $\varepsilon$, depending
linearly on momentum $p$ ($\varepsilon = v_{F}p$), also results in
the cyclotron motion but with the frequency
$\omega_{C}=eB/(|\varepsilon|/v_F^2)$, in which one easily
identifies the energy dependent mass $m=|\varepsilon|/v_F^2$. This
latter equation, equivalent to the Einstein relation between mass
and energy, invokes again the relativistic-like character of
electronic states in graphene. Perhaps surprisingly, the strictly
speaking classical, i.e., linear with the magnetic field,
cyclotron resonance has not been clearly observed in graphene so
far, thought the existing experimental results are well explained
by quantum mechanical approach.

\begin{figure}
\begin{center}
\scalebox{1.25}{\includegraphics*{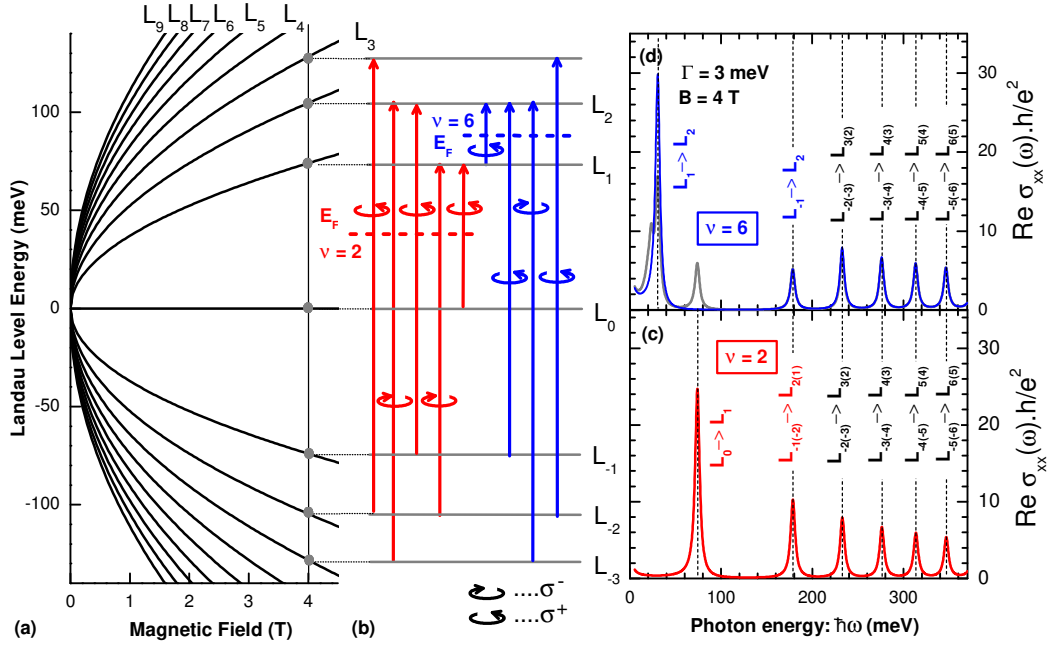}}
\end{center}
\caption{\label{LL_monolayer} (a) A characteristic
$\sqrt{B}$-dependence of LLs in graphene shown for a few low-index
levels (b) Dipole-allowed inter-LL transitions for two different
filling factors $\nu=2$ and 6. Real parts of the (low-temperature)
optical conductivity (proportional to the optical absorption)
determined using Eq.~(\ref{Kubo}) are plotted in panels
(c) and (d) for filling factors $\nu=2$ and 6, respectively. The
grey curve is calculated for the same position of the Fermi level
as in $\nu=6$ case, but at temperature of 150~K.}
\end{figure}

In a quantum-mechanical picture, the application of the magnetic
field $B$ perpendicular to the graphene plane transforms the
continuous electronic spectrum into discrete and highly degenerate
Landau levels (LLs)~\cite{ZhengPRB02}:
\begin{equation}\label{LL_graphene}
E_n=\mathrm{sign}(n)v_F\sqrt{2|e|\hbar B
|n|}=\mathrm{sign}(n)E_1\sqrt{|n|},\quad n=0,\pm1,\pm2 \ldots
\end{equation}
whose positions are defined by a single material parameter, the
Fermi velocity $v_F$ ($E_1=v_F\sqrt{2\hbar e B}$). The degeneracy
of each Landau level is $\zeta(B)=g_vg_s |eB|/h$, where we take
into account both spin and valley degeneracies. This LL spectrum
consists of electron levels ($n>0$), hole levels ($n<0$) and the
zero LL ($n=0$) which is shared by both types of carriers and
which is responsible for the unusual sequence of the quantum Hall
effect in graphene~\cite{NovoselovNature05,ZhangNature05}. We also
immediately see that LLs in graphene are non-equidistant and they
evolve as $\sqrt{B}$, see Fig.~\ref{LL_monolayer}a, what both can
be understood as a consequence of the extreme non-parabolicity (in
fact linearity) of the bands. The unusual $\sqrt{B}$-dependence of
LLs in graphene is responsible for its surprising sensitivity to
the magnetic field. Experimentally, the well-defined LLs have been
spotted in this system down to 1~mT and almost up to temperature
of liquid nitrogen~\cite{NeugebauerPRL09}. It might be realistic
that Landau level quantization in pure graphene could also be
observable in the magnetic field of the Earth
($B_{\mathrm{Earth}}\sim10^{-5}$~T), which is unique for a
condensed-matter system.

Interaction of light with graphene in a quantising magnetic field
has been explored several times both theoretically and
experimentally over the few past
years~\cite{PeresPRB06,GusyninPRB06,SadowskiPRL06,GusyninPRL07,JiangPRL07,GusyninJoPCM07,SadowskiSSC07,SadowskiIJMPB07,DeaconPRB07,PlochockaPRL08,OrlitaPRL08II,NeugebauerPRL09,HenriksenPRL10}.
To describe the magneto-optical response of graphene, we start
with a single-particle model and use the Kubo-Greenwood
formula~\cite{KuboJPSJ57,GreenwoodPPS58,MoseleyAJP78}. Then, the
longitudinal conductivity takes the following form:
\begin{equation}\label{Kubo}
\mathrm{Re}(\sigma^{\pm}(\omega,B))=\frac{4e^2}{\omega}\frac{|eB|}{h}\frac{\gamma}{\pi}
\sum_{m,n}|\langle
m|\hat{v}_{\pm}|n\rangle|^2\frac{f_n-f_m}{(E_m-E_n-\hbar\omega)^2+\gamma^2},
\end{equation}
where $0 \leq f_{n} \leq 1$ stands for the occupation of the
$n$-th LL and $\gamma$ is a phenomenological broadening
parameter. The matrix elements of the velocity operators
$\hat{v}_{+}=(\hat{v}_{x}-i\hat{v}_{y})/\sqrt{2}$ and
$\hat{v}_{-}=(\hat{v}_{x}+i\hat{v}_{y})/\sqrt{2}$ are $|\langle
m|\hat{v}_{+}|n\rangle|^2=\alpha\delta_{|m|,|n|+1}$ and $|\langle
m|\hat{v}_{-}|n\rangle|^2=\alpha\delta_{|m|,|n|-1}$, respectively,
with $\alpha=v_F^2/2$ if $n$ or $m$ equals zero, and otherwise
$\alpha=v_F^2/4$~\cite{SadowskiPRL06,SadowskiIJMPB07,GusyninJoPCM07,AbergelPRB07}.

We see that graphene exhibits a relatively rich (multi-mode)
magneto-optical response, where energies of individual resonances
which correspond to individual inter-LL transitions scale as
$\sqrt{B}$, preserving this unique property of the Landau level fan
chart in graphene. Let us now discuss a few basic facts which
imply from Eq.~(\ref{Kubo}): All dipole-allowed inter-LL
transitions in graphene follow the selection rules $|n|
\rightarrow |n|+1$ and $|n| \rightarrow |n|-1$, which are active
in the $\sigma^{+}$ and $\sigma^{-}$ polarization of the incoming
light \cite{SadowskiIJMPB07,AbergelPRB07}, respectively. The
possible transitions can be divided into three groups ($j\geq 1$):
Inter-band resonances $L_{-j} \rightarrow L_{j+1}$ and $L_{-j-1}
\rightarrow L_{j}$ at energy
$E_1\left(\sqrt{j+1}+\sqrt{j}\right)$, intra-band resonances $L_j
\rightarrow L_{j+1}$ and $L_{-j-1} \rightarrow L_{-j}$ with energy
$E_1\left(\sqrt{j+1}-\sqrt{j}\right)$ and the mixed
L$_{-1(0)}\rightarrow$~L$_{0(1)}$ resonance, involving the $n=0$
LL, having energy of $E_1$. From a purely terminological point of view,
only intra-band resonances can be referred to as cyclotron
resonance (CR), nevertheless, the term \emph{Cyclotron resonance
experiments} is used frequently to describe any of inter- or
intra-band inter-LL transitions in the current literature, and
becomes thus equivalent to a more general term of \emph{Landau
level spectroscopy}.

A scheme of optically active transitions for a hypothetical system
of 2D Dirac fermions at two, arbitrary chosen, filling factors
$\nu=2$ and 6 is illustrated in Fig.~\ref{LL_monolayer}b. The
corresponding optical-conductivity spectra are presented in
Figs.~\ref{LL_monolayer}c and \ref{LL_monolayer}d. The intra-band
transitions appear at low energies and are followed by inter-band
resonances at higher energies. There is however no distinct
separation between these two types of transitions, what is in
contrast to the case of conventional 2D systems based on gapped
semiconductors, but somehow similar to the case of narrow-gap
II/VI compounds structures~\cite{SchultzJPCM98,SchultzPRB98}.
Nevertheless, in graphene, we deal with only one type of atomic
orbital, and therefore both intra- and inter-band transitions
follow similar selection rules: modulus of the LL index is changed
by 1. This is again in contrast to the case of convectional 2D
system, made for instance of GaAs, for which the inter-band
transitions conserve the LL index and their dipole moment is due
to different $s$- and $p$-orbitals of the conduction and valence
band, respectively. Owing to the electron-hole symmetry of the
graphene band structure, two different inter-band resonances, such
as, for example $L_{-2} \rightarrow L_{3}$ and $L_{-3} \rightarrow
L_{2}$ in Fig.~\ref{LL_monolayer}b, may appear at the same energy.
Such degenerated in energy transitions are however active in
opposite circular polarization of light. At low temperatures one
may expect at most two different intra-band transitions but a
series of inter-band transitions. The situation is even more
complex at higher temperatures, when the thermal spreading of the
Fermi distribution exceeds the separation between Landau levels.
The intra-band absorption (CR) may then also reveal a
multi-mode character, as seen, for example, in
Fig.~\ref{LL_monolayer}d, where the grey curve depicts the thermal
activation of L$_{0}\rightarrow$L$_1$ and L$_{2}\rightarrow$L$_3$
transitions. Such multi-mode intra-band absorption spectrum, which
envelop corresponds, nota bene, to the classical cyclotron resonance
discussed at the beginning of this section, was recently observed
in graphene (on the surface of bulk graphite) by Neugebauer
\emph{et al.}~\cite{NeugebauerPRL09}.

\begin{figure}
\begin{center}
\scalebox{0.80}{\includegraphics*{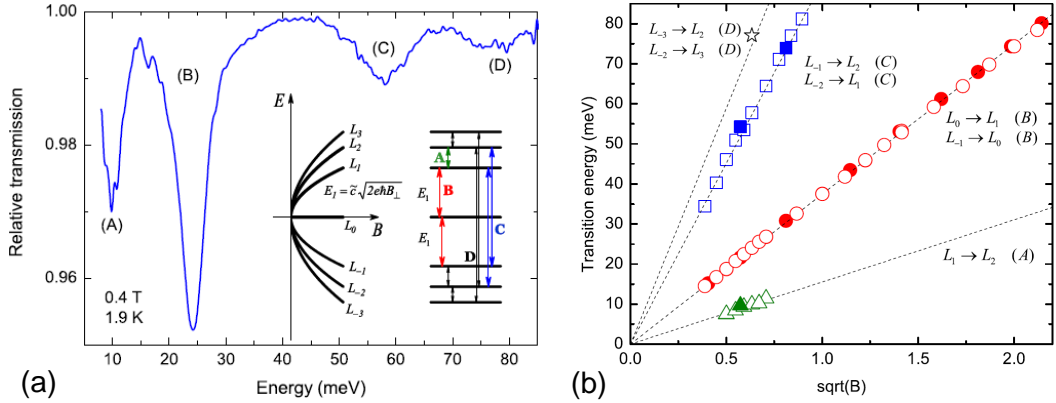}}
\end{center}
\caption{\label{Sadowski} (a) Typical multi-mode
magneto-transmission spectrum of graphene, taken on a few-layer
epitaxial specimen (3-5 layers) prepared on C-terminated surface
of 4H-SiC~\cite{SadowskiPRL06}. The inset shows a schematic of the
evolution of Landau levels with applied magnetic field, and
possible optical transitions. (b) Fan chart of observed inter-LL
transitions, showing clear $\sqrt{B}$ evolution of all absorption
lines. Reprinted from~\cite{SadowskiPRL06}, copyright (2006) by
The American Physical Society.}
\end{figure}

Another interesting point is that the velocity operators (and also
their matrix elements) are magnetic-field-independent in graphene,
contrary to the case of standard systems with a parabolic
band~\cite{AndoJPSJ75}. In consequence, if the occupation
difference $(f_{n}-f_{m})$ between the initial and final state
Landau levels is not changed with the magnetic field (which is
often the case of inter-band transitions), the oscillator strength
of such a transition rises as $\sqrt{B}$, see Eq.~(\ref{Kubo})
with $\omega\sim \sqrt{B}$. The transitions involving the $n=0$
Landau level are also interesting in this respect. In the quantum
limit, once the Fermi energy is pinned to the $n=0$ level, the
oscillator strength of the degenerate (non-polarized)
L$_{-1(0)}\rightarrow$~L$_{0(1)}$ transition does not depend on
the particular occupation $f_{0}$ of this level. This
is due to the fact that the change in one of the occupation
factors ($f_{-1}-f_{0}$) and ($f_{0}-f_{1}$) relevant for the
L$_{-1}\rightarrow$~L$_{0}$ and L$_{0}\rightarrow$~L$_{1}$
transitions, respectively, is compensated by the same gain for the
other one. For the same reason, the oscillator strength of the
unpolarized transition involving the $n=0$ LL increases as a
$\sqrt{B}$ at high magnetic fields. This is very much in contrast
to the case of conventional electron gas with parabolic dispersion
relations, for which the oscillator strength of the cyclotron
resonance is just a measure of the electron concentration,
including the high-field limit. However, the doping of graphene in
this limit can still be deduced from the comparison of the
intensities of the (polarized) L$_{0}\rightarrow$~L$_{1}$ and
L$_{-1}\rightarrow$~L$_{0}$ transitions.

\begin{table}
\caption{\label{Table_graphene_velocity} Fermi velocity in various
graphene specimens deduced by different experimental techniques:
magneto-optics (MO), ARPES, and STS. Values for nearest-neighbour
coupling parameter $\gamma_0$ were obtained via relation $\gamma_0=2\hbar
v_F/(\sqrt{3}a_0)$, $\gamma_0 [\mathrm{eV}] \doteq 3.088 v_F
[10^6\mathrm{m.s}^{-1}]$.}
\begin{indented}
\item[]
\begin{tabular}{@{}llll}
\br
Graphene specimen& $v_F$~(10$^6$~m.s$^{-1}$)& $\gamma_0$~(eV) &Remark\\
\br
Exfoliated, SiO$_2$ substrate          &1.12     &3.46     &MO~\cite{JiangPRL07}\\
Exfoliated, SiO$_2$ substrate          &1.09     &3.37     &MO~\cite{DeaconPRB07}\\
Epitaxial (C-terminated surface)          &1.02-1.03&3.15-3.18&MO~\cite{SadowskiPRL06,OrlitaPRL08II}\\
Epitaxial (C-terminated surface)          &1.13     &3.49     &STS~\cite{MillerScience09}\\
Epitaxial (C-terminated surface)          &1.0      &3.1      &ARPES~\cite{SprinklePRL09}\\
Decoupled on surface of bulk graphite  &1.00     &3.09     &MO~\cite{NeugebauerPRL09}\\
Decoupled on surface of bulk graphite  &0.79     &2.4      &STS~\cite{LiPRL09II}\\
\br
\end{tabular}
\end{indented}
\end{table}

\begin{figure}
\begin{center}
\scalebox{0.40}{\includegraphics*{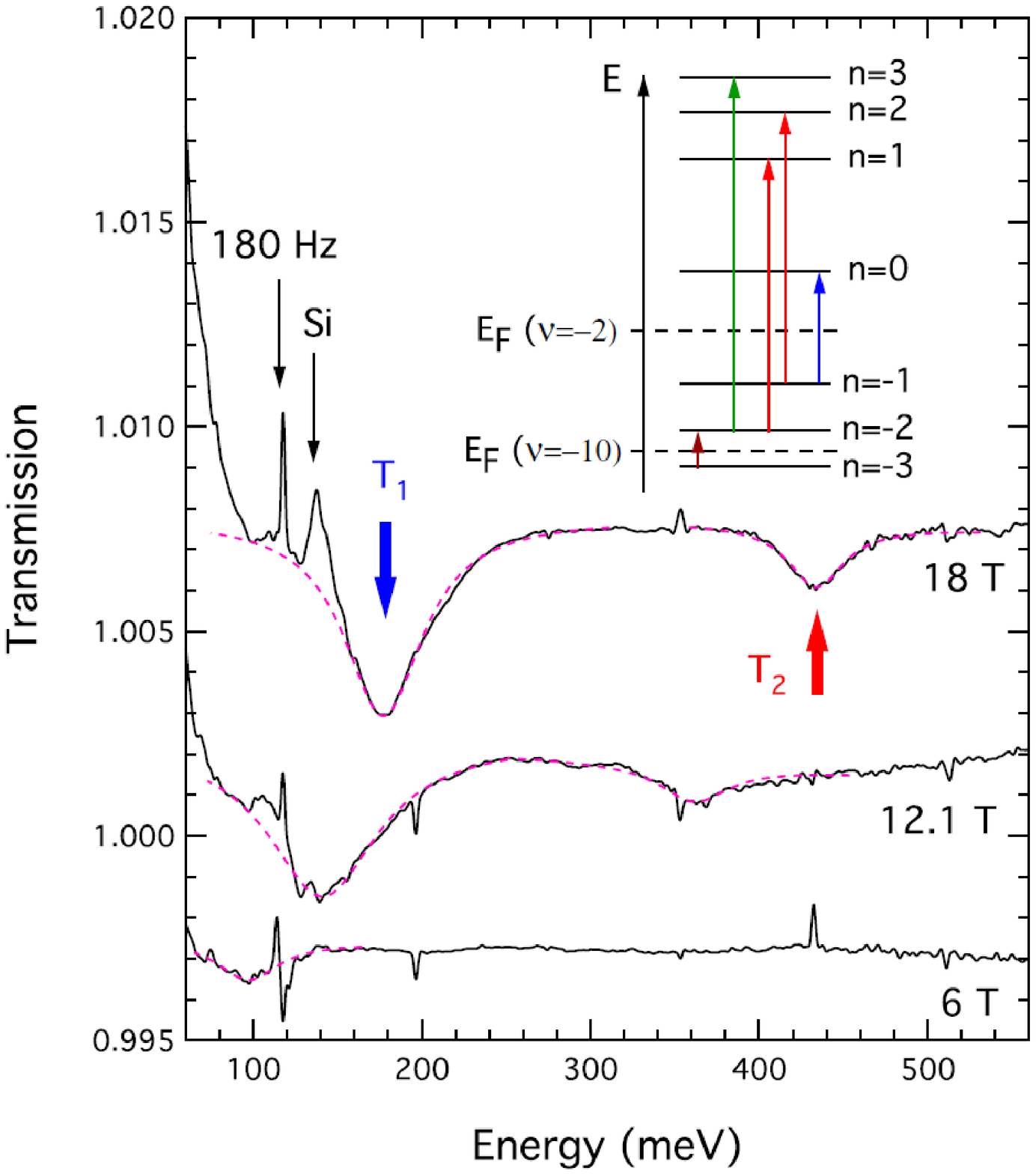}} \hspace{0.5cm}
\scalebox{0.35}{\includegraphics*{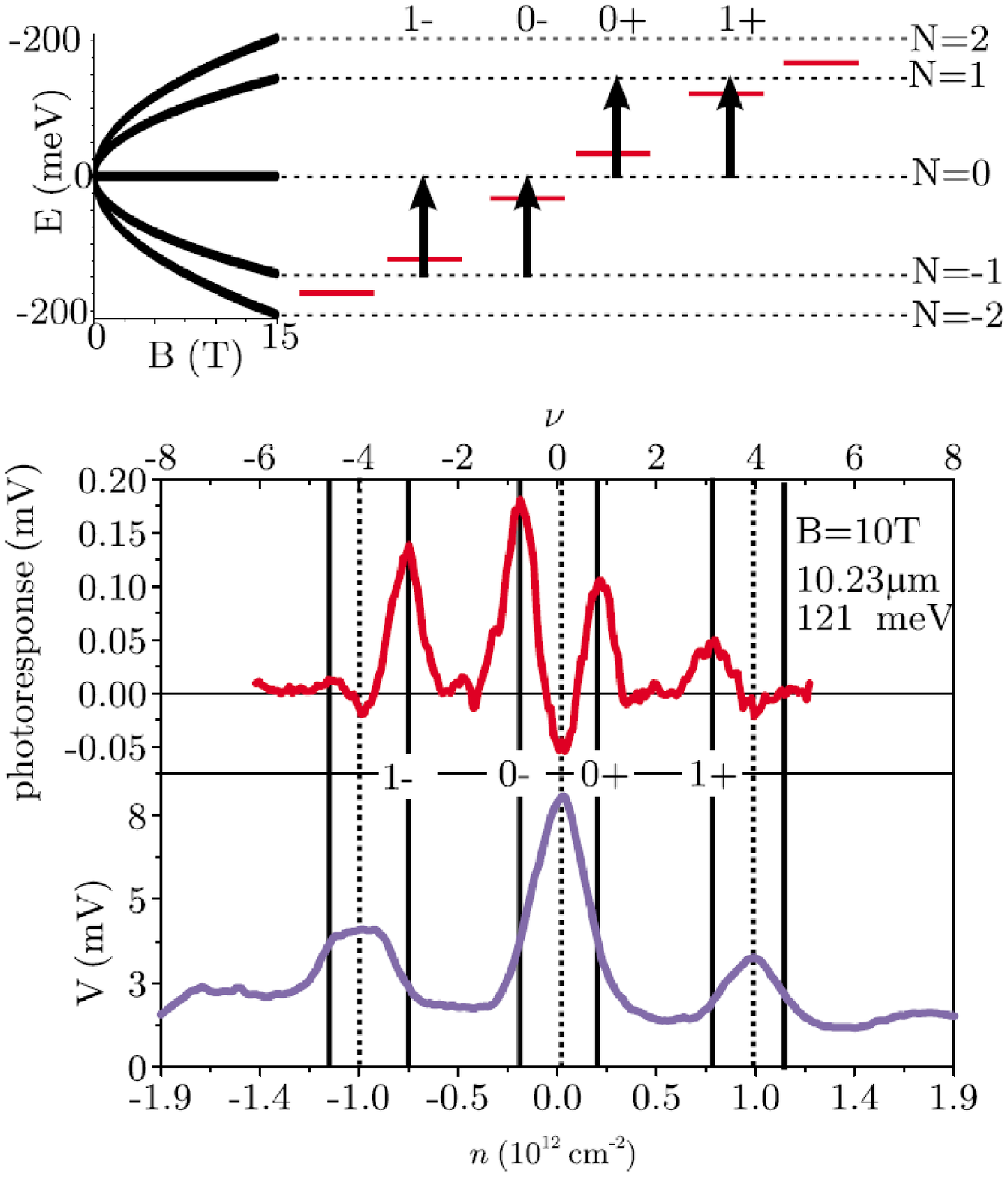}}
\end{center}
\caption{\label{CR_in_Exfoliated_Graphene} (a)
Magneto-transmission spectra of exfoliated graphene on Si/SiO$_2$
substrate obtained by Jiang~\emph{et al.}~\cite{JiangPRL07}.
Individual inter-LL transitions are identified in the inset. (b)
Density dependence of the two-contact resistive voltage and
photoconductive response of an exfoliated graphene specimen at the
fixed magnetic field and fixed energy of CO$_2$ laser, for details
see Deacon \emph{et al.}~\cite{DeaconPRB07}. Reprinted
from~\cite{JiangPRL07,DeaconPRB07}, copyright (2007) by The
American Physical Society.}
\end{figure}

\emph{Experiments:} Magneto-optical measurements, mainly
magneto-transmission, have been up to now performed on three
different kinds of graphene samples. First experiments were
carried out on epitaxial graphene prepared by the thermal
decomposition of the surface of silicon
carbide~\cite{SadowskiPRL06,SadowskiSSC07}, see
Fig.~\ref{Sadowski}. To be more specific, these CR data have been
obtained on a multilayer epitaxial graphene (MEG) specimen
prepared on C-terminated surface of 4H-SiC. The Dirac-like
spectrum, genuine of the graphene monolayer, found to dominate in MEG
structures, has been initially an intriguing observation but this
is today a well-established experimental
fact~\cite{MillerScience09,SprinklePRL09}. The electronic bands in
MEG grown on the carbon face of SiC are alike those of a single
layer because of preferentially rotational and not Bernal-type
layer stacking in this material~\cite{HassPRL08,SprinklePRL09}. To
our knowledge, no reliable magneto-optical measurements have been
reported on samples grown on Si-face of 4H-SiC (or 6H-SiC), where
mostly the standard (well-known from bulk graphite) Bernal
stacking is set. On the other hand, Si-face samples have been used
in a number of ARPES experiments, see
e.g.~\cite{OhtaScience06,BostwickNaturePhys07,ZhouNatureMater07}.
The magneto-optical experiments on epitaxial samples were soon followed by
measurements on a single flake of exfoliated
graphene~\cite{JiangPRL07,DeaconPRB07}, see
Fig.~\ref{CR_in_Exfoliated_Graphene}, and recently, also by CR in
decoupled graphene sheets on the surface of bulk
graphite~\cite{NeugebauerPRL09}. To compare, whereas the
large-size samples (up to a few mm$^2$) covered with MEG flakes
(up to a few mm$^2$) allow for the relatively easy and precise CR
measurements on well-defined, practically undoped graphene sheets
($n_0\approx5\times10^9$~cm$^{-2}$~\cite{OrlitaPRL08II}), the
demanding experiments on single flakes of exfoliated graphene are
more subtle. They have been successfully realized with a
differential (modulation) technique when using the gate voltage to
tune the Fermi energy and therefore to block the apparent
transitions in the reference signal. The magneto-transmission
experiments offer several pieces of important information about
electronic properties of graphene, we will summarize them in the
following.

\emph{Landau level fan charts and Fermi velocity:} A clear
illustration of the characteristic $\sqrt{|Bn|}$ scaling of Landau
levels, in fact equivalent to the observation of linear dispersion
relations of carriers, has been the first important feat of the
Landau level spectroscopy of graphene
systems~\cite{SadowskiPRL06}. Sadowski \emph{et
al.}~\cite{SadowskiPRL06} observed a practically perfectly
$\sqrt{|Bn|}$-scaled Landau level fan chart in MEG structures and
extracted $v_F=1.03\times 10^6$~m.s$^{-1}$ for the Fermi velocity
in this system. Jiang \emph{et al.}~\cite{JiangPRL07} and
subsequently Deacon \emph{et al.}~\cite{DeaconPRB07} found somehow
higher values $v_F\approx1.1\times10^6$~m.s$^{-1}$ for gated
(exfoliated) graphene flakes on Si/SiO$_{2}$. We note that Landau
levels in MEG structures as well as in graphene flakes floating on
the graphite surface have been recently also visualized using
tunnelling spectroscopy in magnetic
fields~\cite{LiPRL09II,MillerScience09}. The Fermi velocity found
in these later STS experiments agrees well with
magneto-transmission data in case of MEG
structures~\cite{SadowskiPRL06,OrlitaPRL08II} but it is
surprisingly low $0.79\times10^6$~m.s$^{-1}$~\cite{LiPRL09II} in
case of graphene flakes on graphite, for which the
magneto-spectroscopy gives much higher value of
$1.00\times10^6$~m.s$^{-1}$~\cite{NeugebauerPRL09}. Individual
values of the Fermi velocity determined in various samples by
different experimental techniques have been summarized in
Tab.~\ref{Table_graphene_velocity}.

\emph{Beyond simple band models:} Relatively small but noticeably
deviations of bands from their ideal linearity, of the order of a
few percent at large ~$\pm0.5$~eV distance from the Dirac point,
have been found by a combination of far and near-infrared
magneto-optical experiments performed on multi-layer epitaxial
graphene by Plochocka~\emph{et al.}~\cite{PlochockaPRL08}. These
deviations were revealed by a departure of the observed
transitions from a simple $\sqrt{B}$-dependence, which rises with
the photon energy of the probing light. No signs of the
electron-hole asymmetry has been found in these experiments. On
the other hand, the traces of the electron-hole asymmetry have
been reported by Deacon~\emph{et al.}~\cite{DeaconPRB07} in
exfoliated graphene placed on Si/SiO$_{2}$ substrate, who
estimated the difference in the electron and hole Fermi velocities
to be of the order of few percents. Magneto-transmission
experiments if carried out on neutral graphene specimens may also
bring a relevant information on a conceivable appearance of a gap
at the Dirac point. Working in the limit of low magnetic fields,
Orlita \emph{et al.} have estimated a gap to be smaller than 1~meV
in quasi-neutral MEG structures~\cite{OrlitaPRL08II} and its
maximum possible value of a fraction of 1~meV in graphene flakes
on graphite substrates~\cite{NeugebauerPRL09}.

\begin{figure}
\begin{center}
\scalebox{0.45}{\includegraphics*{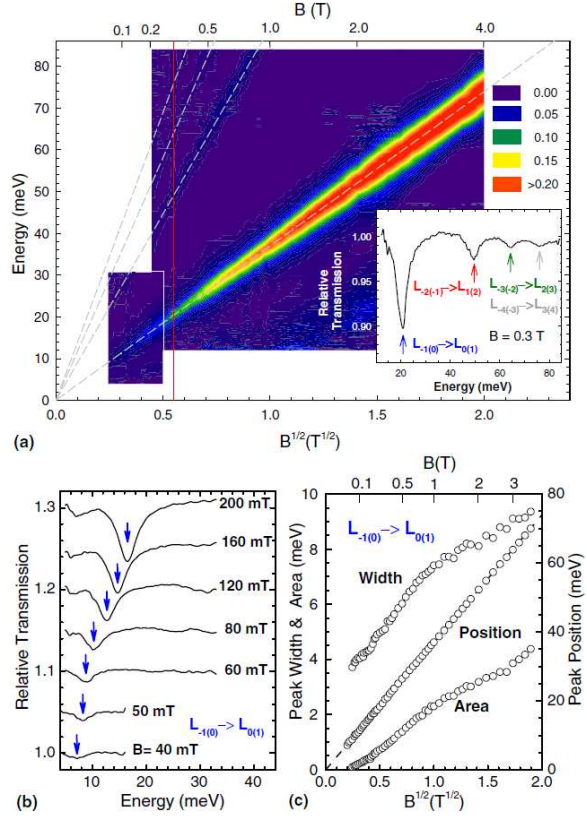}}
\end{center}
\caption{\label{OrlitaLowField} Part (a): Far infrared
transmission $\mathcal{T}$ plotted as $-\ln{\mathcal{T}}$  as a
function of the magnetic field at $T=2.0$~K. The dashed lines
denote the expected transitions for
$v_F=1.02\times10^{6}$~m.s$^{-1}$. The inset shows the
transmission spectrum at $B=0.3$~T. Part (b): FIR transmission
measured at $T=2$~K and low magnetic fields. Successive spectra
are shifted vertically by 0.05. The part (c) shows the peak
position, width and area for the L$_{-1(0)}\rightarrow$L$_{0(1)}$
transition. The dashed line in part (c) is a least squares fit to
the peak positions. Reprinted from~\cite{OrlitaPRL08II}, copyright (2008) by
The American Physical Society.}
\end{figure}

\emph{Scattering/disorder :} Cyclotron resonance measurements on
graphene, in particular in the limit of low magnetic fields (low
frequencies) can be effectively used to estimate the scattering
time and/or mobility of carriers. For instance, Orlita~\emph{et
al.}~\cite{OrlitaPRL08II} (working in fields down to 10~mT range)
have shown (see Fig.~\ref{OrlitaLowField}) the possibility to achieve the room-temperature carrier
mobility exceeding 250 000~cm$^2$/(V.s), which is the record value
among all other known materials. Recent predictions based on the
analysis of transport data taken on exfoliated
graphene~\cite{MorozovPRL08} have been thus confirmed. Moreover,
the closer look at the CR lineshape can offer valuable information
about the scattering mechanisms. An increase of the
(LL-independent) linewidth nearly following the
$\sqrt{B}$-dependence indicates that the short-range scattering
dominates in this particular case~\cite{ShonJPSJ98}. Using also
the magneto-optical methods, the mobility of charged carriers
exceeding $10^7$~cm$^2$/(V.s) up to the temperature of liquid
nitrogen has been determined for high-quality graphene flakes on
the surface of bulk graphite~\cite{NeugebauerPRL09}. CR
measurements thus became a straightforward method to characterize
the quality of graphene specimens, see, e.g.,
Ref.~\cite{JerniganNL09}.

\emph{Electron-electron interaction:} Since the discovery of
graphene, the effects of electron-electron interaction were a
subject of particular interest in this material. Nevertheless, a
great majority of experimental results obtained on various
graphene systems are fairly well understood within single particle
models. This also concerns a number of magneto-transmission
studies~\cite{SadowskiPRL06,PlochockaPRL08,OrlitaPRL08II,NeugebauerPRL09}.
Characteristically, they display a regular, defined by a single
parameter $v_{F}$ series of transitions which are thus very
tempting to be assigned as those between single particle Landau
levels. However, the excitations between highly degenerate Landau
levels are known as nontrivial processes which involve the effect
of electron-electron interaction. Such an electron-hole
excitation is characterized by its wavevector (nota bene
proportional to electron-hole separation). The specific shapes of
dispersion relations of inter and intra Landau level excitations
are central for the many-body physics of the integer
\cite{KlitzingPRL80,NovoselovNature05,ZhangNature05} and
fractional~\cite{TsuiPRL82,DuNature09,BolotinNature09} quantum
Hall effects, respectively. We know from this physics that, when
considering a single parabolic band of a conventional 2DEG (with
equidistant Landau levels), the energies of optically active $k=0$
inter Landau level excitations correspond to those of single
particle excitations. This can be viewed as a consequence of Kohn
(or Larmor) theorem or seen as a result of perfect cancellation of
Coulomb binding and a positive exchange term for the $k=0$
electron-hole excitation. This reasoning does not hold for a 2D
gas of Dirac electrons, for which the exchange term may even
largely exceed the Coulomb binding, both in addition expected to
be different for different pairs of Landau levels. The apparent
approximate validity of the Kohn theorem in graphene is a
surprising effect, and in our opinion calls for further
clarifications on the theoretical background. The first
theoretical works dealing with this problem have been already
published~\cite{IyengarPRB07,BychkovPRB08,AsanoEP2DS18}. We note
however here that although small but noticeable deviations from a
perfect single particle scaling of inter Landau level excitations
have been already reported in experiments on exfoliated graphene
structures~\cite{JiangPRL07}. Very recently, Henriksen~\emph{et
al.}~\cite{HenriksenPRL10} have reported changes in the energy of
the L$_{-1(0)}\rightarrow$~L$_{0(1)}$ transition, especially
pronounced at high magnetic fields, when tuning the Fermi energy
in between $n=-1$ and $n=1$  Landau levels. Both these
observations~\cite{JiangPRL07, HenriksenPRL10} are discussed in
terms of electron-electron interactions but perhaps also include
some effects of disorder. Notably, magneto-optics allows for
probing the nature of quasi neutral graphene in high magnetic
fields (a possible appearance of a gap in the zero LL at filling
factor zero), which recently became a subject of many theoretical
considerations and experimental
works~\cite{KatsnelsonMT07,GiesbersPRL07}.

\section{Bilayer graphene}
\label{DOUBLE}

\begin{figure}
\begin{center}
\scalebox{0.4}{\includegraphics*{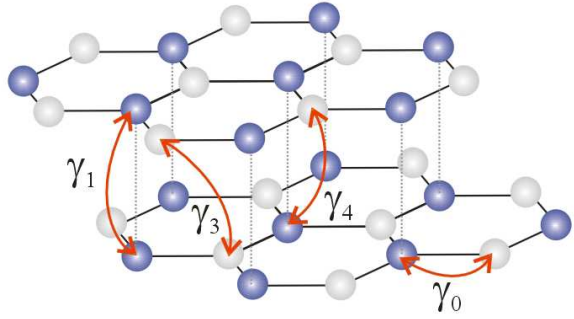}}
\end{center}
\caption{\label{Bilayer_atomic} Crystal structure of the Bernal-stacked
graphene bilayer with the corresponding SWM hopping parameters
$\gamma_0,\gamma_1,\gamma_3$ and $\gamma_4$.}
\end{figure}

\subsection{Band structure}

The interest in the physics of  a graphene bilayer started after
magneto-transport experiments of Novoselov~\emph{et
al.}~\cite{NovoselovNaturePhys06} who showed a characteristic
quantum Hall effect and a Berry's phase of $2\pi$ with
magneto-resistance measurements of this material. Carriers in a
graphene bilayer are characterized by a finite mass but are
clearly distinct from those in conventional systems with parabolic
bands and referred to as \emph{massive Dirac fermions}. Similarly
to the case of graphene, the basics of the electronic band
structure of the graphene bilayer have been established in 1947 by
Wallace~\cite{WallacePR47} and further developed in the band
structure model of bulk graphite introduced by Slonczewski, Weiss
and McClure~\cite{SlonczewskiPR58,McClurePR57,McClurePR60} (SWM
model). As a matter of fact, it is not graphene, but the graphene
bilayer, which is the basic unit in the construction of
Bernal-stacked bulk graphite.

The presence of four atoms in the unit cell of the graphene
bilayer implies the appearance of four bands in the vicinity of
the Fermi level, instead of two in the graphene monolayer. The SWM
model, reduced to a true bilayer, represents a solid basis for
their description. This model implies one intra-layer $\gamma_0$
and three inter-layer $\gamma_1,\gamma_3,\gamma_4$ hopping
integrals, see Fig.~\ref{Bilayer_atomic}. In addition, the
difference in on-site energies of stacked and unstacked triangular
sublattices $\Delta'$, as well as, the bias voltage $U$ applied
across the layers are usually taken into account when modelling
the band structure of the graphene bilayer.

\begin{figure}
\begin{center}
\scalebox{1.2}{\includegraphics*{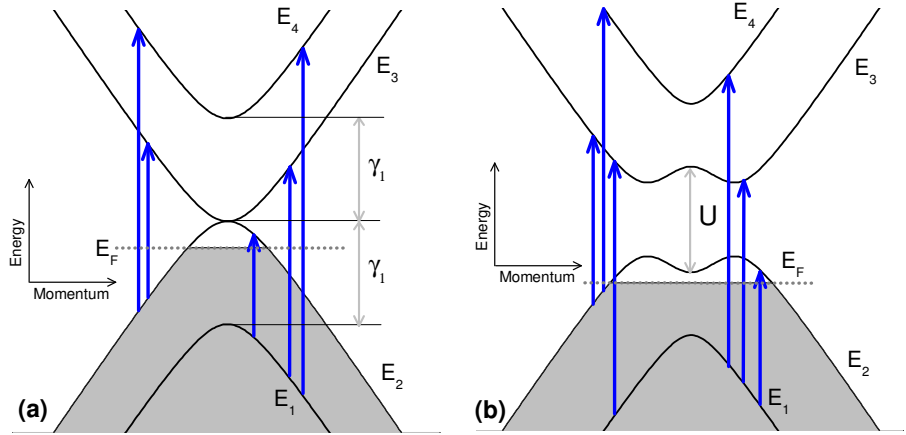}}
\end{center}
\caption{\label{Bilayer_transitions} A schematic view of the band
structure in graphene bilayer, showing its modification via bias
voltage $U$ applied across the two layers, (a) $U=0$ and (b) $U\neq 0$. The arrows show
dipole-allowed transitions in this system which determine its
optical response in the infrared spectral range.}
\end{figure}

As only $\gamma_0$, $\gamma_1$ and $U$ parameters predominantly
affect the shape of bands, we neglect the others for a while and
express the Hamiltonian of the graphene bilayer
as~\cite{McCannPRL06}:
\begin{equation}\label{Bilayer}
\hat{H}=\left(\begin{array}{cccc}
U/2 & 0 &0& v_F\pi^\dag\\
0 &  -U/2 & v_F\pi &0\\
0 & v_F\pi^\dag &-U/2&\gamma_1\\
v_F\pi & 0 &\gamma_1&U/2
\end{array} \right),
\end{equation}
which corresponds to the case of atoms in layers 1 and 2 ordered
as $(A_1,B_2,A_2,B_1)$. The electronic bands obtained by the
diagonalization of this Hamiltonian have simple form of:
\begin{eqnarray}\label{Bilayer_bands}
 \lefteqn{E_{1,2}=-E_{4,3}=}\nonumber \\
\hspace{1cm}
=-\left(\gamma_1^2/2+U^2/4+v_F^2p^2\pm\sqrt{\gamma_1^4/4+v_F^2p^2(\gamma_1^2+U^2)}\right)^{1/2}.
\end{eqnarray}
These bands are symmetric with respect to the zero energy level,
defined by the touching point of the bands $E_2$ and $E_3$
(so-called charged neutrality point) for $U=0$, see
Fig.~\ref{Bilayer_transitions}a. A nearly parabolic shape of bands
$E_2$ and $E_3$ close to the charge neutrality point allows us to
introduce an effective mass $m=\gamma_1/(2v_F^2)$ with a
relatively low value of $m\approx0.03m_0$~\cite{HenriksenPRL08}
comparable to a mass found in narrow gap semiconductors. Assuming
no doping or external gating, the Fermi level is located just at
the charge neutrality point and similarly to graphene, the
graphene bilayer can be viewed as a zero-gap semiconductor
(semimetal). As sketched in Fig~\ref{Bilayer_transitions}b, the
external bias $U$ applied across the layers, transforms this
gap-less band structure into a system characterised by a finite
energy gap $E_{\mathrm{gap}}\approx U$.

\subsection{Optical conductivity at $B=0$}

\begin{figure}
\begin{center}
\scalebox{0.35}{\includegraphics*{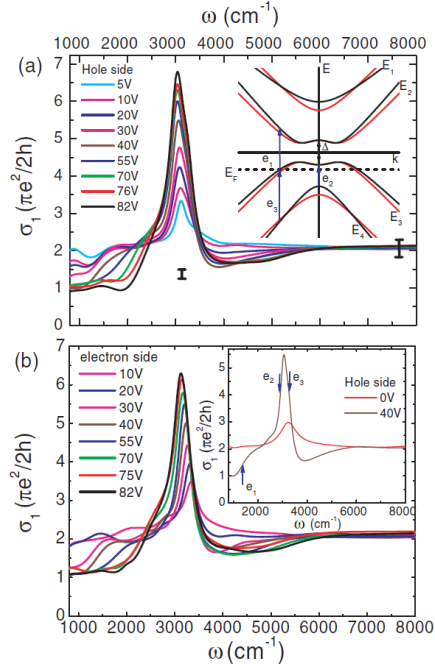}}
\end{center}
\caption{\label{Li_bilayer} Real part of the optical conductivity
of the bilayer graphene as a function of the gate voltages, i.e.
at various hole (a) and electron (b) densities measured by
Li~\emph{et al.}~\cite{ZQLiPRL09}. The inset in part (a)
schematically shows the band structure of the graphene bilayer
without and with parameters $\gamma_4$
and $\Delta'$ taken into account. Position of individual
band-edge inter-band transitions in the optical conductivity is shown in the
inset of the part (b). Reprinted from~\cite{ZQLiPRL09}, copyright (2009) by
The American Physical Society.}
\end{figure}

The relevant optical response of the graphene bilayer is expected
in the infrared range, see the illustrative
Fig.~\ref{Bilayer_transitions}, as the characteristic band energy
scale is given by the parameter
$\gamma_1\approx0.4$~eV~\cite{Brandt88}. Experiments have followed
the well established theoretical
background~\cite{AbergelPRB07,McCannSSC07,NicolPRB08,NilssonPRB08,BenfattoPRB08}.
Two groups almost simultaneously reported on experiments performed
on a single gated flakes of the exfoliated graphene
bilayer~\cite{ZhangPRB08,ZQLiPRL09,KuzmenkoPRB09,KuzmenkoPRB09II},
see Fig.~\ref{Li_bilayer}, overall confirming the expected
electronic band structure given by Eq.~(\ref{Bilayer_bands}). A more closer
data analysis revealed a weak electron-hole asymmetry, going
beyond the Hamiltonian~(\ref{Bilayer}) and quantified by
parameters $\gamma_4\approx150$~meV and
$\Delta\approx18$~meV~\cite{ZhangPRB08}, see also
Tab.~\ref{Table_graphene_bilayerA}. No pronounced influence of the
trigonal warping, expressed predominantly by the $\gamma_3$
parameter, was found in these experiments. Very recently,
analogous experiments have been performed on more sophisticated
devices with two gates incorporated, allowing independent tuning
of the carrier density and voltage $U$~\cite{ZhangNature09,MakPRL09}.
A possibility to smoothly tune the energy gap from zero up to
$E_{\mathrm{gap}}\approx 200$~meV has thus been demonstrated. The
possibility of relatively easily tuning of the gap in the graphene
bilayer depicts the interesting potential of this system in
designing the electronic devices.

\begin{figure}
\begin{center}
\scalebox{1.25}{\includegraphics*{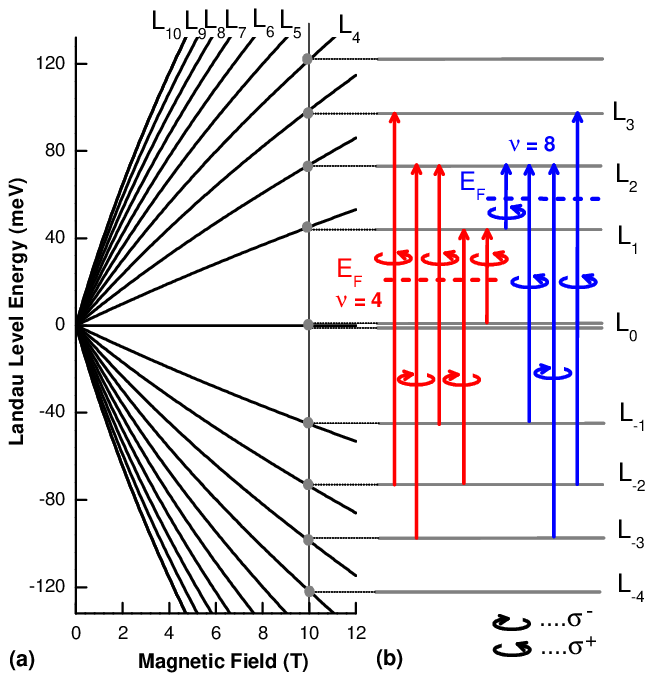}}
\end{center}
\caption{\label{LL_bilayer} a) LLs in graphene bilayer, evolving
nearly linearly with $B$ at lower fields, but bending to
sub-linear dependence at higher $B$ and/or higher energies. (b)
Dipole-allowed inter-LL transitions for the case of two filling
factors $\nu=4$ and 8.}
\end{figure}

\begin{table}
\caption{\label{Table_graphene_bilayerA} Band structure parameters
of the graphene bilayer $\gamma_1, \gamma_4$ and $\Delta'$ derived
from infrared optical experiments performed on the exfoliated
graphene bilayer specimen.}
\begin{indented}
\item[]
\begin{tabular}{@{}llll}
\br
Band structure parameter&  Value & References\\
\br
$\gamma_1$                 &400 meV          &\cite{ZhangPRB08,ZQLiPRL09}\\
                           &378 meV          &\cite{KuzmenkoPRB09,KuzmenkoPRB09II}\\
                           &$\approx350$ meV &\cite{HenriksenPRL08}\\
\mr
$\gamma_4$                 &150 meV          &\cite{ZhangPRB08,ZQLiPRL09}\\
                           &140 meV          &\cite{KuzmenkoPRB09,KuzmenkoPRB09II}\\
\mr
$\Delta'$                  &18 meV          &\cite{ZhangPRB08,ZQLiPRL09}\\
                           &22 meV          &\cite{KuzmenkoPRB09,KuzmenkoPRB09II}\\
\br
\end{tabular}
\end{indented}

\caption{\label{Table_graphene_bilayerB} Various quantities
related to the graphene bilayer in the parabolic approximation of
$E_2$ and $E_3$ bands. The numerical values are calculated for
$\gamma_1$=375~meV and $v_F=10^6$~m.s$^{-1}$. $m_0$ denotes the
bare electron mass. }
\begin{indented}
\item[]
\begin{tabular}{@{}llrl}
\br
Quantity &Relations& Numerical values& \\
\br
Effective mass $m$ & $\gamma_1/(2v_F^2)$ & 0.033&$m_0$\\
Density of states  &  $g_s g_v \gamma_1/(4v_F^2\pi\hbar^2)$ & $2.76\times10^{10}$ & cm$^{-2}$.meV$^{-1}$\\
Carrier density  &  $g_s g_v \gamma_1|E_F|/(4v_F^2\pi\hbar^2)$ & $2.76\times10^{10}|E_F|$ & cm$^{-2}$\\
Cyclotron energy $\hbar\omega_c$ & $2 \hbar eB v_F^2/\gamma_1$ & $3.51B[T]$ & meV\\
\br
\end{tabular}
\end{indented}
\end{table}

\subsection{Magneto-spectroscopy}

In most of the experimentally studied cases, the
graphene bilayer evokes the quantum mechanical character of its band structure
under the magnetic field applied across the 2D plane. Following the Hamiltonian
(\ref{Bilayer}), the energy ladder of Landau levels with indices
$n\geq0$, is given by:
\begin{eqnarray}\label{LL_bilayer_general}
 \lefteqn{E^n_{1,2}=-E^n_{4,3}=}\nonumber \\
\hspace{1cm} =-\left(\gamma_1^2/2+(n+1/2)E_1^2\pm\sqrt{\gamma_1^4/4
+(n+1/2)E_1^2\gamma_1^2+E_1^4/4}\right)^{1/2},
\end{eqnarray}
The eightfold ($8eB/h$) degeneracy of the LL with $n=0$ is
twice the degeneracy of all other levels and this results in a
characteristic quantum Hall effect with the Berry phase of
$2\pi$~\cite{NovoselovNaturePhys06,McCannPRL06}. At energies
around the neutrality point, the Landau levels are formed of $E_2$
and $E_3$ bands, and their energies are practically linear with
$B$. As shown in Fig.~\ref{LL_bilayer}a, a departure from
(saturation of) this linear dependence appears however at higher
energy distances from the neutrality point (from the zero LL at
energy zero) and/or for Landau levels with higher indices.

Similarly to the case of the graphene monolayer, the magneto-optical
response of graphene bilayer can be described within the
Kubo-Greenwood formalism, employed in the previous Chapter, but
the individual matrix elements are not easy to evaluate. Up to
now, this problem was approached approximatively by
Mucha-Kruczy\'nski \emph{et al.}~\cite{MuchaKruczynskiJPCM09} and
also some numerical results are
available~\cite{KoshinoPRB08,KoshinoSSC09}. In general, we expect
that all transitions between LLs arising from all four bands are
dipole-allowed, if the selection rules $n\rightarrow n\pm1$ are
obeyed, and they are active in the $\sigma^{\pm}$ polarization of
the incoming light, respectively. A few transitions in the
vicinity of the Fermi level, including their polarization,  are
schematically shown in Fig.~\ref{LL_bilayer}b, for two arbitrarily
chosen filling factors $v=4$ and 8. Certain additional
complications arise from the appearance of the trigonal warping,
expressed mainly by the parameter $\gamma_3$, which is not
included in the Hamiltonian (\ref{Bilayer}) and which induces the
second set of dipole-allowed transitions between LLs. Such
transitions appearing between levels whose indices differ by $3N\pm1$, where
$N=1,2,3\ldots$~\cite{AbergelPRB07}, may have their oscillator
strengths comparable with $n\rightarrow n\pm1$ transitions but only
at low magnetic fields~\cite{AbergelPRB07}.

The scheme of inter Landau level transitions in a graphene bilayer
is largely simplified when a parabolic approximation of $E_2$ and
$E_3$ bands is considered~\cite{McCannPRL06}, i.e. when the
Hamiltonian~(\ref{Bilayer}) is reduced to 2$\times$2 matrix (the
$E_1$ and $E_4$ split-of bands are neglected). This approach seems
to be fairly sufficient when considering the low-energy
excitations in an undoped or weakly doped graphene bilayer. The LL
spectrum then takes the form of
$E^n_3=-E^n_2=\hbar\omega_c\sqrt{n(n+1)}$ for $n\geq0$, or
equivalently:
\begin{equation}\label{LL_bilayer_parabolic}
E_n=\mathrm{sign}(n)\hbar\omega_c\sqrt{|n|(|n|+1)},\quad n=0,\pm1,\pm2\ldots,
\end{equation}
where $\omega_c$ stands for the cyclotron frequency
$\omega_c=eB/m$ with $m=\gamma_1/(2v_F^2)$. In this approximation,
the Landau levels are perfectly linear with the applied field
$B$. Let us note that in the limit of high $|n|$ (practically even for
$|n|\geq1$), the LL spectrum Eq.~(\ref{LL_bilayer_parabolic}) has
the form $E_n\approx\mathrm{sign}(n)\hbar\omega_c(|n|+1/2)$,
typical of conventional massive particles.

Within the parabolic approximation, the matrix elements of the velocity
operators, $|\langle m|\hat{v}_{\pm}|n\rangle|^2$, to be set in
the Kubo-Greenwood formula, have been evaluated by Abergel and
Fal'ko~\cite{AbergelPRB07}. They obtained: $|\langle
m|\hat{v}_{\pm}|n\rangle|^2=l_B\omega_c\left(\sqrt{|n|+1}\right)\delta_{|m|,|n|\pm1}$ (for $m\neq0$ and $n\neq0$), where $l_B=\sqrt{\hbar/(eB)}$
denotes the magnetic length. Transitions involving doubly
degenerated $n=0$ LL level must be taken into separately. The
spectrum is obviously expected to depend on the actual position of
the Fermi energy and special care has to be taken to account for
possible splitting of the $n=0$ Landau level (including the
elucidation of the character of this splitting). To prevent
confusion, we note that our convention of LL indexing differs from
the one used by Abergel and Fal'ko, whose Landau level spectrum
reads: $E^n_3=-E^n_2=\hbar\omega_c\sqrt{n(n-1)}$ for $n\geq 1$, as compared to
our Eq.~(\ref{LL_bilayer_parabolic}).

\begin{figure}
\begin{center}
\scalebox{0.35}{\includegraphics*{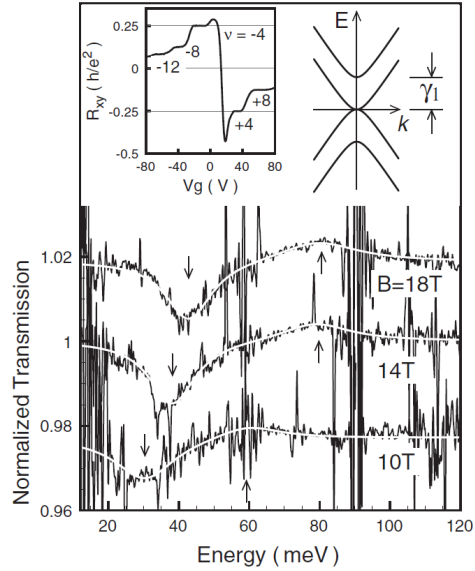}}
\end{center}
\caption{\label{Henriksen} Magneto-transmission spectrum of an exfoliated
graphene bilayer measured by Henriksen~\emph{et al.}~\cite{HenriksenPRL08}.
Upper left inset: the quantum Hall effect of the bilayer graphene sample measured in
situ at $B=18$~T. Upper right inset: schematic of the zero-field dispersion
relation of bilayer graphene. Reprinted from~\cite{HenriksenPRL08}, copyright (2008) by
The American Physical Society.}
\end{figure}

The available magneto-optical data on a graphene bilayer are not
as rich as on graphene. Probably the only report is by Henriksen
\emph{et al.}~\cite{HenriksenPRL08}, who examined a single gated
flake of an exfoliated graphene bilayer in (differential) far
infrared magneto-transmission experiments. In this
work~\cite{HenriksenPRL08} the authors investigated the
transitions between the adjacent LLs (intra-band transitions) as a
function of the magnetic field and the carrier density (filling
factor). A relatively good agreement with the simplified LL energy
spectrum~(\ref{LL_bilayer_general}) has been obtained for low carrier
densities. At higher filling factors, deviations appear due to the
finite energy gap induced by the gate voltage, as recently
explained by~Mucha-Kruczy\'{n}ski \emph{et
al.}~\cite{MuchaKruczynskiSSC09}.

\section{Graphite}
\label{BULK}

\subsection{Band structure}

\begin{figure}
\begin{center}
\scalebox{0.6}{\includegraphics*{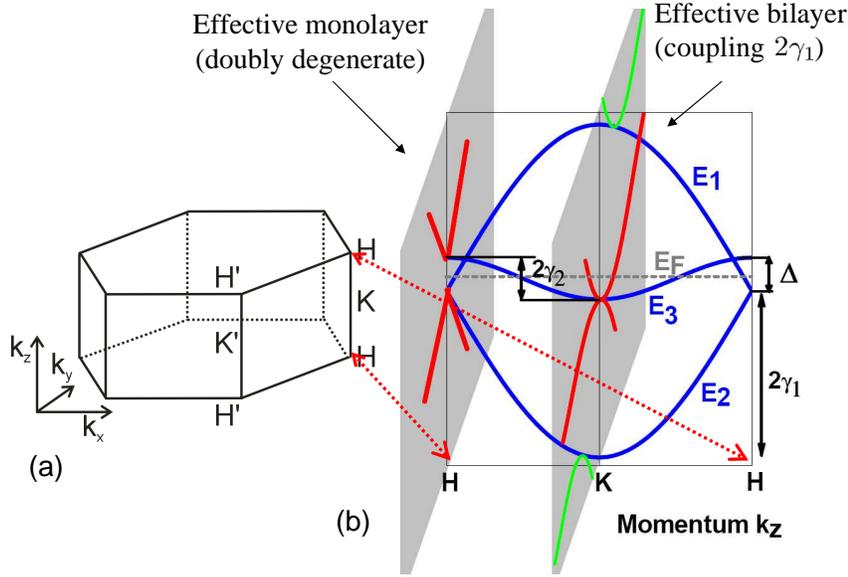}}
\end{center}
\caption{\label{Graphite_band_structure} Brillouin zone of
graphite and its electronic band structure along the $H$-$K$-$H$
line. Whereas electrons in the vicinity of the $K$ point behave as
massive Dirac fermions with mass twice enhanced in comparison to
the true graphene bilayer, the holes around the $H$ point have
nearly linear dispersion and behave as (nearly) massless Dirac
fermions in graphene, but with an additional double degeneracy.}
\end{figure}

The renewed interest in the properties of bulk graphite is a
direct consequence of the outbreak of the current graphene
physics. As a 3D crystal, graphite is a system characterised by a
higher degree of complexity compared to graphene, nevertheless,
both material share many common properties.

The physics of bulk graphite has been reviewed several times, see
e.g. Refs. ~\cite{ChungJMS02,Brandt88}. Here we mostly focus on
optical properties which distinctly uncover the massless and
massive Dirac-like electronic states also in this material. The
appealing possibility to trace the ``relativistic'' carriers not
only in graphene monolayer and bilayer, but also in bulk graphite
(which is definitely easier to handle!) resulted in a number of
works which offer new pieces of information, new interpretations
of old data but often also rediscoveries of well-established
knowledge.

\begin{figure}
\begin{center}
\scalebox{0.4}{\includegraphics*{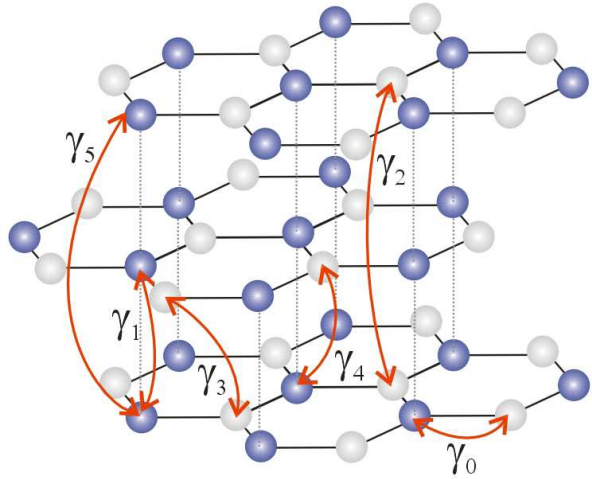}}
\end{center}
\caption{\label{Bulk_atomic} Segment of the crystal structure of
A-B stacked graphite with coupling constants used in the standard
SWM model.}
\end{figure}

Starting with the pioneering work of Wallace~\cite{WallacePR47}, the
fundamentals of the graphite band structure have been formulated by
Slonczewski, Weiss and McClure in
fifties~\cite{McClurePR57,SlonczewskiPR58,McClurePR60}, as already mentioned in
Section~\ref{DOUBLE}. This model describes the band structure along the
$H$-$K$-$H$ line of the 3D Brillouin zone, see
Fig.~\ref{Graphite_band_structure}, which is responsible for most of the
electrical and optical properties of bulk graphite. The SWM model mostly
implies six hopping integrals $\gamma_0,\ldots,\gamma_5$, as sketched in
Fig.~\ref{Bulk_atomic}, and additional parameter $\Delta$, usually referred to
as pseudogap and related to the difference of the crystal field on atom-sites A
and B. By definition, this pseudogap is related to a similar parameter in an
isolated graphene bilayer as $\Delta'=\Delta-\gamma_2+\gamma_5$.

To describe optical response of graphite, we leave the full
complexity of the SWM model and follow a simplified approach
taking account only two most relevant hopping integrals
$\gamma_0$ and $\gamma_1$, which describe intra- and interlayer
tunnelling, respectively \cite{PartoensPRB07,KoshinoPRB08}. In
this way, the band structure of bulk graphite along the
$H$-$K$-$H$ is obtained by the diagonalization of the Hamiltonian:
\begin{equation}\label{EffectiveBilayer}
\hat{H}=\left(\begin{array}{cccc}
0 & 0 &0& v_F\pi^\dag\\
0 &  0 & v_F\pi &0\\
0 & v_F\pi^\dag & 0&\lambda\gamma_1\\
v_F\pi & 0 &\lambda\gamma_1& 0
\end{array} \right),
\end{equation}
which is formally equivalent to that of an unbiased graphene
bilayer~(\ref{Bilayer}), with the same ordering of the atomic
wavefunctions, but with an effective coupling tuned by the
momentum $k_z$ in the direction perpendicular to layers,
$\lambda=2\cos(\pi k_z)$. The appearance of the cosine band in
$k_z$, having the amplitude of $2\gamma_1$, thus reflects the
periodic crystal ordering along the $c$-axis, similarly to
formation of minibands in semiconductor
superlattices~\cite{EsakiIBM70}.

The straightforward diagonalization of the
Hamiltonian~(\ref{EffectiveBilayer}) gives the band structure in
the form of~Eq.~(\ref{Bilayer_bands}), i.e. electronic
bands equivalent to the graphene bilayer, but with an effective
coupling $\lambda\gamma_1$:
\begin{equation}
\label{EffectiveBilayer_bands}
E_{1,2}=-E_{4,3}=-\left((\lambda\gamma_1)^2/2+v_F^2p^2\pm\sqrt{(\lambda\gamma_1)^4/4+v_F^2p^2(\lambda\gamma_1)^2}\right)^{1/2}.
\end{equation}
Hence, depending on their momentum along the $c$-axis, electrons
in bulk graphite behave as massive Dirac fermions in graphene
bilayer, but with an effective coupling $\lambda\gamma_1$, which
directly gives their effective mass $m=\lambda\gamma_1/(2v_F^2)$.
For example, electrons at the $K$ point ($k_z=0$) behave as
massive Dirac fermions with an effective mass enhanced twice
($\lambda=2$) in comparison to true graphene bilayer. Carriers (in
real graphite holes) at the $H$ point ($k_z=0.5$) have a character
of massless Dirac fermions due to the effectively vanishing
inter-layer coupling ($\lambda=0$). Bulk graphite thus shares some
common properties with the graphene monolayer as well as its
bilayer.

\begin{figure}
\begin{center}
\scalebox{0.35}{\includegraphics*{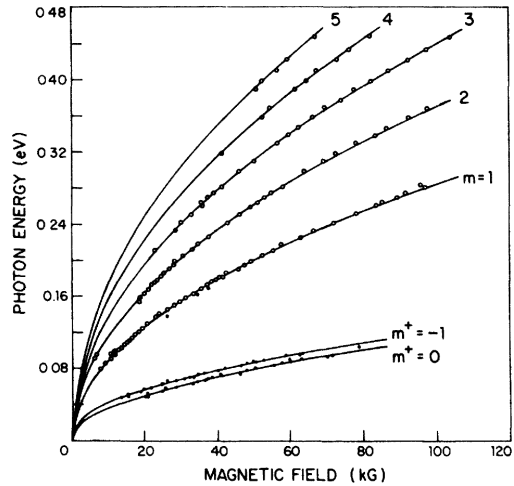}}
\end{center}
\caption{\label{Toy} Positions of inter-LL transitions related to
the $H$ point of bulk graphite observed by Toy~\emph{et
al.}~\cite{ToyPRB77} in magneto-reflection experiment, see Fig.~\ref{Toy}. The
observed well-defined $\sqrt{B}$-dependence is characteristic of
massless Dirac fermions. The splitting of the line with the lowest
energy is induced by existence of the pseudogap $\Delta$. Reprinted from~\cite{ToyPRB77}, copyright (1977) by
The American Physical Society.}
\end{figure}

\subsection{Optical conductivity at $B=0$}

In reflection and/or transmission experiments at zero magnetic
field the whole optical response corresponds to an average over
all momenta $k_z$ along the $H$-$K$-$H$ line. With this averaging,
the concluded response (per one sheet) is surprisingly close to
the optical conductivity of an isolated graphene monolayer, as
pointed out by Kuzmenko~\emph{et al.}~\cite{KuzmenkoPRL08}, see
also Refs.~\cite{LiPRB06,LiPRB09}. Hence, a nearly universal
optical conductivity can also be observed in bulk graphite.

\subsection{Magneto-spectroscopy}

\begin{figure}
\begin{center}
\scalebox{0.75}{\includegraphics*{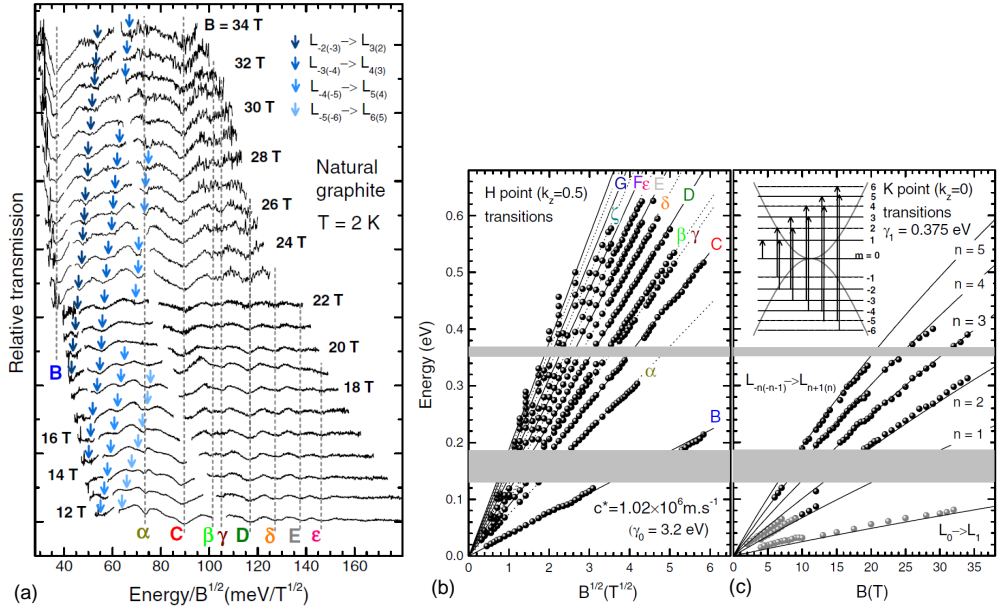}}
\end{center}
\caption{\label{Orlita_graphite} (a) Transmission spectra of a thin graphite
layer at selected magnetic fields. The plotted energy is scaled as $E/\sqrt{B}$
to emphasize the massless Dirac fermion-like features originating at the $H$
point (indicated by dashed vertical lines). Arrows denote transitions arising
at the $K$ point which evolve (nearly) linearly with $B$. (b) Positions of the
absorption lines related to the $H$ point as a function of $\sqrt{B}$. The
solid and dashed lines represent expected positions of absorption lines for
$v_F=1.02\times10^6$~m/s (c) Positions of the absorption lines related to the
$K$ point as a function of $B$. The solid lines show the expected
dipole-allowed transitions in a graphene bilayer with an effective coupling
$2\gamma_1$ calculated using Eq.~(\ref{EffectiveBilayer_bands}) for
$v_F=1.02\times10^6$~m/s and $\gamma_1=375$~meV. The inset schematically shows
the observed inter-band transitions in the effective bilayer. Reprinted
from~\cite{OrlitaPRL09}, copyright (2009) by The American Physical Society.}
\end{figure}

Historically, the verification of the band structure expected in
the framework of the SWM model and estimates of individual
tight-binding parameters have been often done using optical
experiments in magnetic fields. In this case, the main
contribution to the optical response is provided just by the $K$
and $H$ points, where Landau bands become flat, what leads to
singularities in the joint density of
states~\cite{KoshinoPRB08,KoshinoSSC09}. Cyclotron resonance of
massive electrons around the $K$ point was reported in several
works, see older papers
Refs.~\cite{GaltPR56,SchroederPRL68,SuematsuJPSJ72,DoezemaPRB79}
and also recent measurements~\cite{LiPRB06}, showing well-defined
response which scales nearly linearly with $B$. Much less
information has been collected about the massless Dirac fermions
around the $H$ point but nevertheless the features in
magneto-reflection spectra which follow the $\sqrt{B}$-dependence,
typical of 2D massless particles, have been already reported by
Toy~\emph{et al.}~\cite{ToyPRB77}. This work~\cite{ToyPRB77} may
be considered as probably a very first observation of massless
Dirac fermions in solids. Later on, magneto-transmission
experiments on a very thin graphite specimen ($\approx$100 sheets)
allowed for a deeper analysis of the $H$ point optical
response~\cite{OrlitaPRL08,OrlitaJPCM08}. Magneto-optical response
of the $H$ point of graphite includes a set of transitions
equivalent to those of
graphene~\cite{SadowskiPRL06,JiangPRL07,DeaconPRB07}, but it is
nevertheless significantly richer~\cite{OrlitaSSC09}. Since there
are four carbon atoms in the unit cell of graphite (and only two
in the unit cell of graphene) the degeneracy of Landau levels in
graphite (at the $H$ point only) is doubled as compared to the
case of graphene. This results in a set of additional dipole
allowed inter Landau transitions at the $H$ point of graphite
which are indeed observed in experiments, see transitions marked
with small Greek letters in Fig.~\ref{Orlita_graphite}.

The basic understanding of the complex magneto-optical response of
bulk graphite~\cite{OrlitaPRL09}, see Fig.~\ref{Orlita_graphite},
is possible just by viewing this material as an effective graphene
monolayer and an effective bilayer with a coupling strength
enhanced twice in comparison to a true graphene bilayer.
Interestingly, this model including only two material parameters,
$\gamma_0$ and $\gamma_1$, is capable to catch the basic
(magneto-)optical properties of bulk graphite, a complex 3D
material. This creates an interesting link between physics of bulk
graphite and graphene monolayer and bilayer. The values of band
structure parameters determined by magneto-transmission
experiments~\cite{OrlitaPRL09} are $\gamma_0=(3.20\pm0.06)$~eV and
$\gamma_1=(375\pm10)$~meV. The presented model is however
evidently simplified, and therefore, let us now consider the role
of other tight-binding parameters $\gamma_2\ldots\gamma_5$ and
$\Delta$.

The integral $\gamma_2$, due to the hopping between next-nearest
graphene planes (see Fig.~\ref{Bulk_atomic}), is dominantly
responsible for the presence of electron and hole pockets around
$K$ and $H$ points and therefore, i.e., for the semimetallic
character of graphite. This is clearly evidenced for instance by
magneto-transport
experiments~\cite{WoollamPRL70,LukyanchukPRL04,LukyanchukPRL06,SchneiderPRL09}.
Using the language of the effective graphene bilayer and
monolayer, we obtain an effective $n$-doped bilayer at the $K$
point with the Fermi level $E_F\approx20$~meV and similarly,
$p$-doped monolayer around the $H$ point with $E_F\approx-20$~meV.
It is important to note that magneto-optics and magneto-transport
probes are sensitive to very different states selected from those
along the $K$-$H$ dispersion line. Magneto-optics is selective
with respect to the states at $K$ and $H$ points, because it is
sensitive to the maxima in joint density of states (integrated
along the $k_z$ direction) of Landau bands which becomes flat at
the extreme of the Brillouin zone along the $z$ direction. In
turns, magneto-oscillations in transport or de Haas-van Alphen experiments
result from the subsequent coincidences between the Fermi energy position and the maximum in
the density of states. Those latter maxima appear not only at $K$
and $H$ point but importantly also at the intermediate point
somewhere in the middle of the dispersive line. It appears that
$K$- and intermediate-point maxima, both reflecting rather massive
(bilayer) character, are responsible for the two characteristic
sequences of magneto-oscillations observed in magneto-transport
experiments on
graphite~\cite{MikitikPRB06,SchneiderPRL09,SmrckaPRB09}.

Trigonal warping of the band structure of graphite is described in
the SWM model by the $\gamma_3$ parameter ($\gamma_3\approx
300$~meV~\cite{Brandt88}), and its relative influence on the band
structure is scaled with the parameter
$\lambda$~\cite{KoshinoSSC09}. Therefore, the band structure is
mainly modified in the vicinity of the $K$ point and its influence
at the $H$ point is completely cancelled. Probably, the most
pronounced effect related to trigonal warping is the observation
of the harmonics of the basic CR mode due to the $K$-point
electrons. Interestingly, $\gamma_3$ parameter negligibly affects
the LL energies at the $K$ point, but it induces new selection
rules allowing transitions between LLs differing by $3N\pm1$ in
their
indices~\cite{GaltPR56,NozieresPR58,SuematsuJPSJ72,DoezemaPRB79},
whose relative strength is increasing with the decreasing magnetic
field~\cite{AbergelPRB07}. Let us note, however, that
determination of LLs within the full SWM model is only possible
via direct numerical diagonalization of a truncated infinite
matrix~\cite{NakaoJPSJ76}.

The effects of $\gamma_4$ and $\Delta$ parameters are analogous to
the the case of a true bilayer, i.e., they in principle induce the
electron-hole asymmetry of the band structure as discussed in the
previous Section. Again, the effective strength of $\gamma_4$
scales with $\lambda$, giving the most pronounced effects at the
$K$ point, where a relatively weak asymmetry has been
observed~\cite{MisuJPSJ79,MendezPRB80,KuzmenkoPRL08,OrlitaPRL09,ChuangPRB09}.
At the $H$ point, the parameter $\Delta$ corresponds to a
(pseudo)gap opened at the Dirac point in the graphene-like band
structure and is estimated in the range of several
meV~\cite{ToyPRB77,OrlitaPRL08,GruneisPRB08II,OrlitaJPCM08}.

\section{Few-layer Bernal-stacked segments of graphite}
\label{STACKS}

\begin{figure}
\begin{center}
\scalebox{1.3}{\includegraphics*{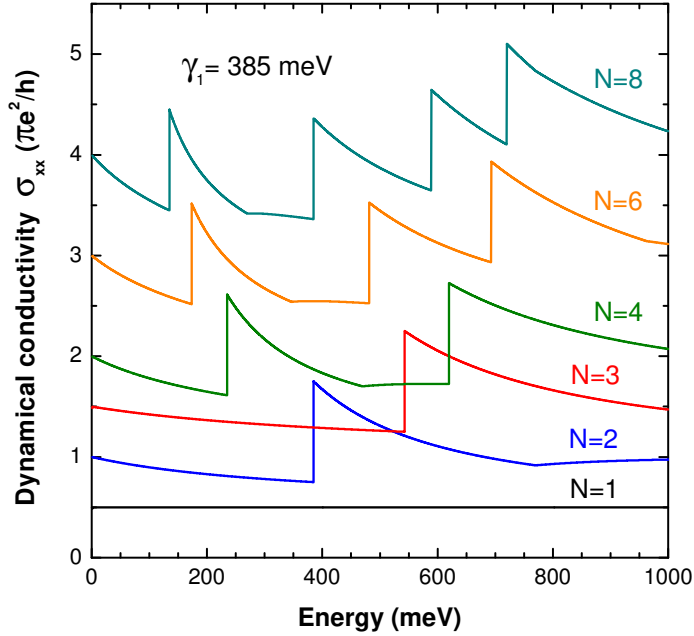}}
\end{center}
\caption{\label{Heinz} The expected dynamic conductivity of Bernal-stacked $N$-graphene layers, as calculated within a
simple band structure model of these systems discussed in the text. All graphene stacks are assumed to be
neutral. The dynamic conductivity of the monolayer is constant and equals $\pi e^2/h$. The bilayer conductivity is
calculated according to the theory developed in Ref.~\cite{AbergelPRB07}, and assuming $\gamma_1 = 385$~meV. The conductivity of the
higher order stack is the sum of the conductivities of its effective bilayers and if applicable of a monolayer (if
$N$ is odd). For example: the trilayer conductivity is the sum of the conductivity of the monolayer and the effective
bilayer with the coupling parameter of $1.41\gamma_1$; the conductivity of the $N=6$ stack is the sum of the conductivities
of three effective bilayers with effective coupling parameters of $0.45\gamma_1$, $1.25\gamma_1$, and $1.80\gamma_1$.}
\end{figure}

A fairly established knowledge about optical properties of the 2D
graphene monolayer and bilayer, as well as 3D bulk graphite
logically leads to a question about the properties of graphene
stacks composed of three and more layers with Bernal ordering of
sheets. As pointed out by Partoens and
Peeters~\cite{PartoensPRB06} the 10-layer graphene stack resembles
already graphite very much, but the investigation of the
graphene-to-graphite evolution is interesting and instructive. To
get insight into the electronic states of a few-layer graphite
stack, we use a relatively simple and intuitive model. We start
with the electronic band structure of bulk graphite and select
discrete values of momentum $k_z$ along the $c$-axis,
$\kappa_m=(\pi/2))[1-m/(N+1)]$, with $m=0,2,\ldots(N-1)$ for odd
$N$ or with $m=1,3,\ldots(N-1)$ for even $N$, for which a standing
wave appears in a stack with $N$ layers. Using a mathematical
language, this approach corresponds to a decomposition of the
Hamiltonian (assumption that $\gamma_2=\gamma_5=0$ is compulsory)
describing a graphite stack with $N$-layers (dimension $2N\times
2N$) into a set of sub-Hamiltonians ($4\times4$). For even $N$,
each segment corresponds to the Hamiltonian of the bilayer with
the effective coupling constant $\lambda_m\gamma_1$, which is
directly related to the allowed discrete values of momentum $k_z$
as $\lambda_m=2\cos\kappa_m$. For odd number of stacked layers
$N$, we get very similar results, $(N-1)/2$ effective bilayers and
one monolayer-like Hamiltonian ($2\times2$). Hence, in this way,
the problem of the band structure of few-layer graphite is reduced
to the already discussed band structures of graphene monolayer and
bilayer, by simple introducing the appropriate effective coupling
parameters. This approach has been discussed by Partoens and
Peeters~\cite{PartoensPRB07} and further developed by Koshino and
Ando~\cite{KoshinoPRB07,KoshinoPRB08,KoshinoSSC09} who also
included the influence of the magnetic field. Perhaps more
elaborated, but less intuitive approach can be found, e.g.,
in~Refs.~\cite{GuineaPRB06,GruneisPRB08II}.

Experimentally, only very first measurements which probe the optical properties of graphene stacks with $N >
2$ have just appeared for stacks up to $N = 8$. They display a clear evolution of the absorption spectra as a
function of $N$ and well resemble the features which can be seen in the traces of the calculated optical
conductivity, expected in these structures (see Fig.~\ref{Heinz}).

\section{Summary and Outlook}

Optical spectroscopy has in recent years been successfully applied
to investigate the properties of new two dimensional allotropes of
carbon and to revise the properties of bulk graphite. Majority of
these experiments have been devoted to establish or verify the
band structure character of the investigated materials and they
certainly helped to establish and/or to better understand the
characteristic nature of massless or massive Dirac electronic
states in
graphene~\cite{LiNaturePhys08,JiangPRL07,DeaconPRB07,NairScience08,MakPRL08},
in graphene-like MEG layers on
SiC~\cite{SadowskiPRL06,PlochockaPRL08,OrlitaPRL08II}, in bilayer
graphene~\cite{ZhangPRB08,ZQLiPRL09,KuzmenkoPRB09,ZhangNature09,MakPRL09},
or in graphite~\cite{OrlitaPRL08,KuzmenkoPRL08,OrlitaPRL09} and
its few layer segments~\cite{MakCM09}. Characteristically for
these semimetalic materials, the methods of infrared spectroscopy
have been frequently used but also (nearly) visible optics
appeared as a relevant tool to uncover their
properties~\cite{NairScience08,PlochockaPRL08,FeiPRB08,GokusACSNano09}.
The authors of this review are particularly attached to
magneto-spectroscopy techniques and consider them among the most
effective methods to studying the band structure parameters.
Estimations of the Fermi velocity in graphene based systems or of
the band curvature (effective mass) in bilayer graphene structures
are the primary results of applications of these methods, see
Tabs.~\ref{Table_graphene_quantities}--\ref{Table_graphene_bilayerB}.
In complement to convectional electric transport measurements, the
magneto-optics is also effective to evaluate the efficiency of
carrier scattering~\cite{OrlitaPRL08II,JerniganNL09} and becomes a
unique method if the electric contacts cannot be applied.

More challenging for experimentalists, such as the authors of this review, is
to uncover the unexpected properties. Identification, with magneto-optical
experiments, of graphene-like bands in MEG structures~\cite{SadowskiPRL06} was
certainly a surprise, but is now a well acknowledged
fact~\cite{HassPRL08,MillerScience09,SprinklePRL09}. Similarly, the
surprisingly extremely weak scattering efficiency anticipated for graphene from
magneto-optical experiments~\cite{NeugebauerPRL09} founds also its
justification in a very recent theoretical work~\cite{BorysenkoCM09}. The
deviations from simple models, tentatively attributed to the effects of
electron-electron interactions, observed in the magneto-optical studies of
graphene~\cite{JiangPRL07,HenriksenPRL10} or in studies at $B=0$~\cite{LiNaturePhys08,MakPRL08} are equally appealing.

It is obviously risky to foresee the further developments in
optics of graphene-related structures but probably safe to say
that they will be very much dependent on the progress in sample
preparation: fabrication of easy handled large area samples, of
systems with higher electronic quality and certainly new
structures. Optical and magneto-optical methods will continue to
be used to extract the relevant band structure parameters of these
materials. These methods are likely to be of particular importance
with respect to intense efforts of band gap engineering in
different graphene systems~\cite{NovoselovNaturePhys06} (e.g.,
graphene bilayer under the applied
voltage~\cite{ZhangNature09,MakPRL09,MuchaKruczynskiSSC09,KuzmenkoPRB09II}
or chemically functionalised
graphene~\cite{EliasScience09,LuoAPL09,LiuAPL09}). The problem of
the specific, for linear bands, role of electron-electron
interactions in inter Landau level transitions is to be clarified
both on the experimental and theoretical ground. Better quality
samples could certainly help in experiments, as they recently did
in the case of the first observations of the fractional quantum
Hall effect in graphene~\cite{DuNature09,BolotinNature09}.
Intriguingly, the classical collective phenomena such as low
energy plasmons are absent in the experiments so far. This is to
be verified, for example, in low-field low-frequency
magneto-optical experiments on samples with well defined shape, as
already done for standard 2D semiconductor
systems~\cite{AllenPRL77,FedorychIJMPB09}. Low-energy plasma
excitations might be very unique in
graphene~\cite{HwangPRB07,PoliniCM09,HillEPL09,TudorovskiyCM09,RoldanPRB09}
and also interesting from the viewpoint of THz
applications~\cite{RyzhiiCM09}. The predicted nonlinear effects
related to cyclotron resonance and/or plasmon excitations are also
to be examined~\cite{MikhailovEPL07,MikhailovPRB09}. A possibility
to detect the cyclotron resonance emission which may perhaps
easily appear in the system with non-equidistant Landau levels,
due expectable inefficient inter-Landau level Auger
processes~\cite{PotemskiPRL91,MorimotoPRB08,PlochockaPRB09}, is
worth to verify. This could give rise to efficient ($B$ tunable)
sources of infrared radiation or even to THz
lasers~\cite{RyzhiiJAP09}. (Magneto-)optical experiments may be
also decisive in confirmation of the Dirac
subbands~\cite{GibertiniPRB09} when fabricating the artificial
graphene~\cite{ParkNL09} (by honeycomb lithography of a
conventional 2DEG). Approaching the illusive quantum
electrodynamics effect of Zitterbewegung is perhaps another
challenge for optics of
graphene~\cite{KatsnelsonEPJB06,RusinPRB07,RusinPRB08,RusinPRB09}.

It should be stressed that this review covers only the basic linear
transmission/reflection optical studies of graphene-based
structures. For example, it does not include the vast domain of
Raman scattering experiments~\cite{FerrariPRL06,MalardPR09}. We just briefly note, that Raman scattering
measurements on graphene systems are so far limited to
investigations of the phonon response. However, since we deal with
resonant processes, the phonon response also includes information
on the electronic bands and, for example, its analysis has become
one of the most popular method of primary characterization of
different graphene structures. The electron-phonon
interaction~\cite{AndoJPSJ06,CatroNetoPRB07} is another important issue that
has been successfully studied in graphene using Raman scattering
techniques~\cite{PisanaNaturMater07,YanPRL07}, including also the application of
magnetic fields~\cite{AndoJPSJ07,GoerbigPRL07,FaugerasPRL09}. Observation of Raman scattering signal
due to electronic excitations in graphene systems~\cite{KashubaPRB09} would be highly
interesting in view of investigations of single particle
versus collective plasmon excitations and/or dispersion relations
(due to electron-electron interaction) of inter Landau level
excitations. Another class of optical experiments, which are not
discussed here, concerns time resolved studies and the related
investigations of carrier dynamics. There are already few reports
in this area~\cite{DawlatyAPL08,GeorgeNL08,SunPRL08,ChoiAPL09,PlochockaPRB09,KampfrathPRL05,BreusingPRL09,NewsonOE09},
which is likely to expand in the future.
Investigations of carrier dynamics with time resolved studies in
the far infrared range and under application of a magnetic field
might be interesting with respect to search for far infrared
emission from graphene structures. One may finally speculate that
these are perhaps the unexpected discoveries, such as the recent
report on broad band emission from graphene in the visible
range~\cite{GokusACSNano09} (induced by oxygen plasma etching), which will open the entirely new domain in optics
of two-dimensional allotropes of carbon.

\section{Acknowledgements}

This work has been supported by the European Commission via EuroMagNET II contract no. 228043 and
Marie Curie Actions MTKD-CT-2005-029671 grant, as well as by PCR CNRS(France)-CNRC(Canada),
CNRS-PICS-4340, MSM0021620834, KAN400100652, GACR no P204/10/1020 and Barrande no 19535NF
projects.

\section{References}
$^{\dag}$ M. O. is also affiliated at the Institute of Physics, Charles
University in Prague, Czech Republic and Institute of Physics ASCR, v.v.i.,
Prague, Czech Republic.\\


\end{document}